\RequirePackage{fix-cm}

\documentclass[twocolumn]{arxiv}

\usepackage{graphicx}

\usepackage[hyphens]{url}
\usepackage{lineno,hyperref}
\usepackage{csquotes} 

\usepackage{rotating}

\usepackage{multirow}

\usepackage{subcaption}

\usepackage{amsmath}

\usepackage{paralist}

\usepackage{fontawesome}

\usepackage[multiple]{footmisc}

\usepackage{threeparttable}

\makeatletter
\let\cl@chapter\undefined
\makeatletter

\usepackage[capitalise,nameinlink]{cleveref}
\crefname{section}{Sect.}{Sect.}
\Crefname{section}{Section}{Sections}
\crefname{figure}{Fig.}{Figs.}
\Crefname{figure}{Figure}{Figures}
\crefname{table}{Tab.}{Tabs.}
\Crefname{table}{Table}{Tables}
\crefname{lstlisting}{List.}{List.}
\Crefname{lstlisting}{Listing}{Listings}

\newcommand{\eg}{e.\,g.,\ }
\newcommand{\ie}{i.\,e.,\ }

\newcommand{\labeltitle}[1]{\vskip 0.03in \noindent\textbf{#1}} 

\journalname{The VLDB Journal}

\usepackage{glossaries}

\newglossaryentry{fpga}{name=FPGA, description={field programmable gate arrays}}
\newglossaryentry{gpu}{name=GPU, description={graphics processing unit}}
\newglossaryentry{rdbms}{name=RDBMS, description={relational database management system}}

\usepackage{breakcites}

\usepackage{mdframed}
\newmdtheoremenv{insight}{Insight}

\usepackage{enumitem}

\begin{document}


\title{Non-Relational Databases on FPGAs: Survey, Design Decisions, Challenges}


 


\author{Jonas Dann \and Daniel Ritter \and Holger Fröning}

\institute{J. Dann \and D. Ritter \at SAP SE, Germany\\ \email{\{firstname.lastname\}@sap.com} \and
    H. Fröning \at Heidelberg University, Germany\\ \email{holger.froening@ziti.uni-heidelberg.de}
}

\date{Received: date / Accepted: date}

\maketitle

\begin{abstract}
Non-relational database systems (NRDS),\linebreak such as graph, document, key-value, and wide-column, have gained much attention in various trending (business) application domains like smart logistics, social network analysis, and medical applications, due to their data model variety and scalability. 
The broad data variety and sheer size of datasets 
pose unique challenges for the system design and runtime (incl. power consumption).
While CPU performance scaling becomes increasingly more difficult, we argue that NRDS can benefit from adding field programmable gate arrays (FPGAs) as accelerators. 
However, FPGA-accelerated NRDS have not been systematically studied, yet. 

To facilitate understanding of this emerging domain, we explore the fit of FPGA acceleration for NRDS with a focus on data model \emph{variety}.
We define the term NRDS class as a group of non-relational database systems supporting the same data model. 
This survey describes and categorizes the inherent differences and non-trivial trade-offs of relevant NRDS classes as well as their commonalities in the context of common design decisions when building such a system with FPGAs. 
For example, we found in the literature that for key-value stores the FPGA should be placed into the system as a smart network interface card (SmartNIC) to benefit from direct access of the FPGA to the network. 
However, more complex data models and processing of other classes (\eg graph and document) commonly require more elaborate near-data or socket accelerator placements where the FPGA respectively has the only or shared access to main memory.
Across the different classes, FPGAs can be used as communication layer or for acceleration of operators and data access.
We close with open research and engineering challenges to outline the future of FPGA-accelerated NRDS.

\end{abstract}



\section{Introduction}
\label{sec:introduction}

Recent business and socio-technical trends like smart applications (\ie leveraging advanced analysis techniques), the internet of things, as well as business and social networks require applications to more efficiently deal with increasingly larger amounts of data in various, non-relational data models close to the underlying domain (cf. \cite{DBLP:journals/corr/abs-1910-09017, journals/csur/DavoudianCL18}).
For example, the predominant data exchange format in distributed business applications is JSON \cite{DBLP:journals/vldb/LangdaleL19}, requiring the processing and storage of nested object (\ie key-value) and array data.
Furthermore, working with JSON documents gains popularity in schema-less contexts that require flexible data models \cite{journals/csur/DavoudianCL18}.
Similarly, applications like social network analysis require processing and storage capabilities of graph data that has a focus on entities and their relationships.  
At the same time the amount of data starkly increases into gigabytes of JSON documents (\eg \cite{DBLP:journals/vldb/LangdaleL19}) and big graph datasets (\eg up to one trillion edges \cite{DBLP:journals/pvldb/ChingEKLM15}).
However, traditional relational database systems fall short on requirements of model \emph{variety} (\ie expressiveness, flexibility) and \emph{efficiency} (\eg data volume, scalability) of these applications \cite{journals/computer/Abadi12, Brewer12, journals/sigmod/Cattell10, journals/cacm/Stonebraker10}.

To address the requirements of variety and efficiency, new classes of systems emerged, namely non-relational database systems (NRDS), that provide a wide variety of database models (\eg graphs, documents) for flexible on-the-fly and application-specific data modeling and efficient processing (\eg by relaxing traditional relational database system constraints).
Similar to recent non-relational data processing surveys \cite{DBLP:journals/corr/abs-1910-09017, journals/csur/DavoudianCL18}, we consider the following data models in this work: graph, document, key-value, and wide-column stores.

\labeltitle{Variety} On first sight, NRDS do not have many commonalities besides coarse-grained system design principles (\eg scalability) and their application specificity.
For example, complex graph query languages are conceptually different from the simple APIs of key-value stores (\eg put, get \cite{conf/sosp/DeCandiaHJKLPSVV07}).
\begin{figure}[bt]
    \centering
    \includegraphics[width=\linewidth]{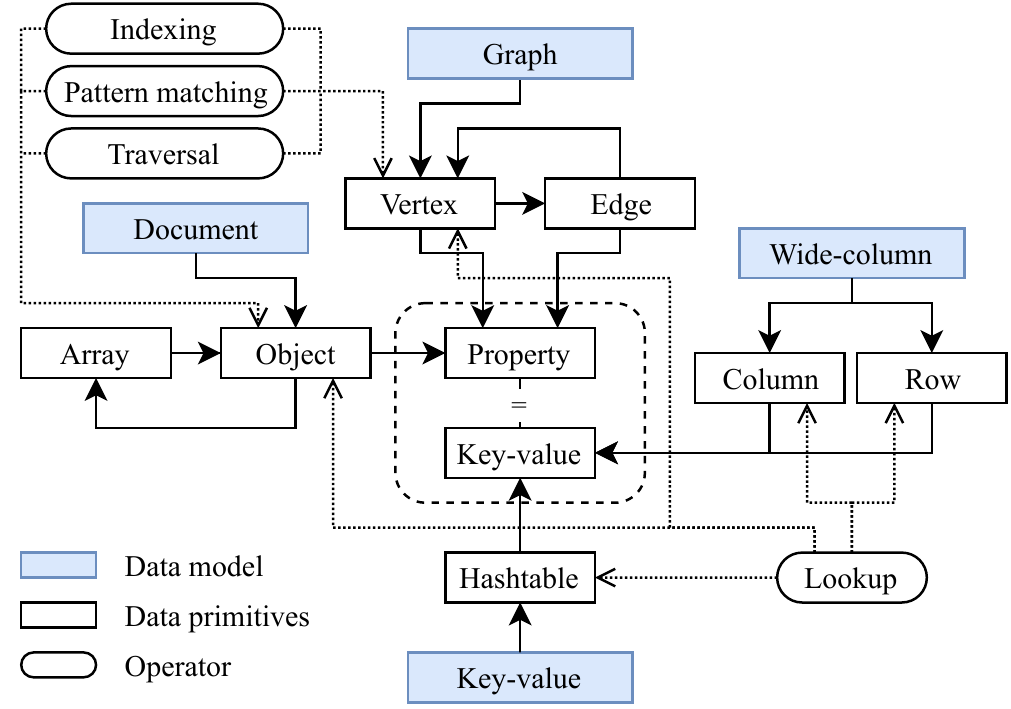}
    \caption{NRDS data models and their relation}
    \label{fig:datamodel}
\end{figure}
Although acknowledging their differences, we argue that distinct non-relational data models share an underlying primitive in key-value pairs and types of operators, like lookup, traversal (\eg BFS on graphs, path traversal on document hierarchies), pattern matching (\eg Cypher \cite{conf/sigmod/FrancisGGLLMPRS18} on graphs, XPath on documents), and indexing (\eg PageRank on graphs, full text document indices).
\Cref{fig:datamodel} depicts those differences and commonalities regarding the data models (blue boxes).
Each model has a differentiating structure layer of specific primitives (\eg vertices and edges of a graph) connecting properties or key-value pairs.
Key-value directly stores key-value pairs in a hash table (\eg \cite{journals/pvldb/IstvanSA17}), while wide-column uses a column name and a row name as key to the key-value pair (\eg \cite{journals/sigops/LakshmanM10}).
Graphs store vertices connected by edges and associate properties with both of those primitives \cite{journals/csur/DavoudianCL18}, while documents are built from objects and arrays that again refer to properties \cite{journals/csur/DavoudianCL18} which are equivalent to key-values.
Furthermore, all data models support simple lookup operations and the more expressive data models (document and graph) support structural traversals, pattern matching, and indexing.

\labeltitle{Efficiency} The NRDS-specific data models and operations are also relevant when considering efficient processing.
For example, graph processing has highly irregular memory-access patterns and little or no temporal and spatial locality.
This is challenging for current NRDS, mostly built on general-purpose CPUs (cf. \cref{sec:background}), due to the CPU's fixed deep cache hierarchy and coarse memory access granularity based on cache line sizes. 
Similar to the application specificity in database systems, there has been a fundamental change in computer architecture and technology trends based on ever rising performance requirements of applications (domain-specific architectures \cite{journals/cacm/HennessyP19}).
Since the end of Dennard scaling \cite{DennardGRBL74}, frequency and therewith performance scaling of CPUs is reaching limits.
Power consumption of computing systems is now a hard constraint leading to energy efficiency of operations such as computations and memory accesses dominating overall performance.
While GPUs as one option offer massive parallelism, they exhibit significantly reduced performance when the internal cores do not execute the same instruction (\ie  warp divergence), \eg common for graphs with varying degrees \cite{journals/csur/ShiZZJHLH18}.

Thus, new computer architectures such as field programmable gate arrays (FPGAs) are explored for future performance scaling in relational database systems and NRDS (\eg \cite{journals/pvldb/IstvanSA17, conf/fccm/OwaidaSKA17, conf/sigmod/SidlerIOKA17}).
FPGAs are reconfigurable computing platforms that can implement custom massively parallel processor architectures. 
There are no structural restrictions on parallelism, all data on an FPGA is bit-addressable, and data types are not restricted to multiples of bytes.
Additionally, FPGAs can be placed anywhere in a system which in particular is important for concepts like near-data processing or processing on the wire which allow for substantial reductions in the amount of data being moved.
As a result, FPGAs are being widely used in various applications, including data centers \cite{conf/isca/PutnamCCCCDEFGGHHHHKLLPPSTXB14}, machine learning \cite{conf/fpga/UmurogluFGBLJV17}, and also data management which is the focus of this survey.
For database systems specifically they might even outperform the currently more widely available GPUs \cite{conf/norchip/RoozmehL17}.
The focus of this survey will thus be FPGA acceleration of NRDS which is not well studied yet.

\subsection{Research gap and related surveys}
\label{sub:related_surveys}
\begin{figure}
    \centering
    \includegraphics[width=0.95\linewidth]{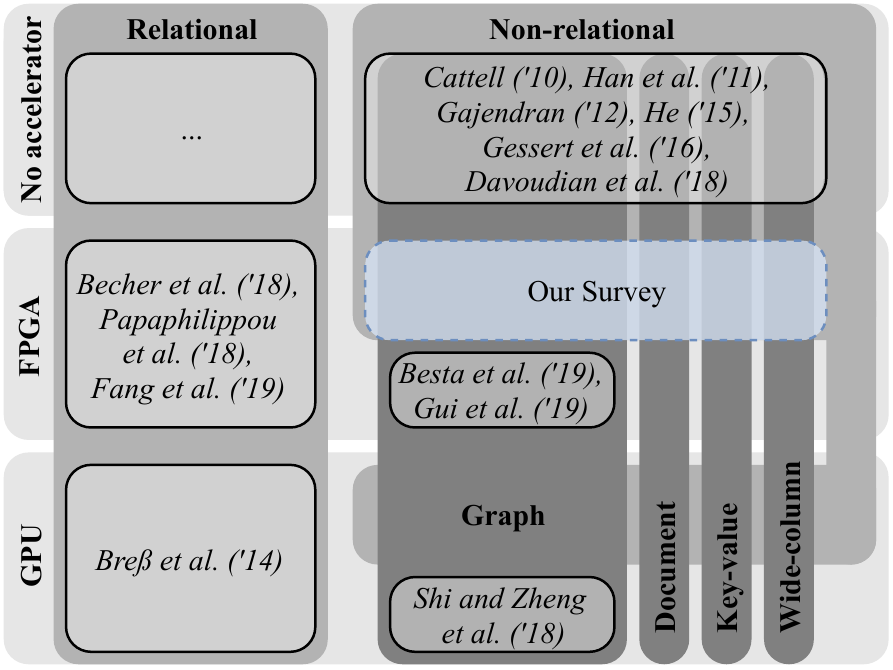}
    \caption{Related surveys and our contribution}
    \label{fig:surveys}
\end{figure}

While FPGA-accelerated NRDS may 
meet the requirements of emerging applications, there is only limited academic work in terms of survey research.
\Cref{fig:surveys} depicts recent surveys related to FPGA-accelerated NRDS.
We consider surveys on \emph{relational} and \emph{non-relational} data in the context of three kinds of acceleration: \emph{no accelerator}, \emph{FPGA}-, and \emph{GPU}-accelerated.
The contribution to non-relational data processing is further specified by the NRDS classes (\ie graph, document, key-value, wide-column).

Leaving the vast amount of work on non-accelerated relational database systems out of scope, there are several related surveys on acceleration of such systems.
In \cite{journals/dbsk/BecherGBDGMPSTW18}, Becker et al.\ pose the challenge of using the on-the-fly reconfigurability of FPGAs in relational databases.
This results in open questions on exploitation of heterogeneous hardware, query partitioning, and dynamic hardware reconfiguration.
Papaphilippou et al.\ categorize the literature into frameworks (\eg Centaur \cite{conf/fccm/OwaidaSKA17} or DoppioDB \cite{conf/sigmod/SidlerIOKA17}) and specialized accelerators for common operators in a relational database, and also highlight upcoming cache coherent connectors for FPGAs (\eg OpenCAPI).
Most recently, Fang et al.\ \cite{journals/vldb/FangMHLH20} state main memory access, programmability, and GPUs as the three biggest factors holding back FPGAs in in-memory relational database systems in the past.
Regarding GPU-accelerated relational database systems, \cite{journals/tlsdkcs/BressHSBS14} raises challenges like the I/O bottleneck and query planning but does not offer convincing solutions.

Non-accelerated NRDS are well-covered by surveys \cite{journals/sigmod/Cattell10, journals/csur/DavoudianCL18, GajendranS12, journals/ife/GessertWFR17, journals/asei/HeC15, conf/pervasive/HanELD11} which are discussed in \cref{sec:background}.
Additionally, there are two surveys covering graph processing as a high-performance computing (HPC) workload \cite{journals/corr/abs-1903-06697, journals/jcst/Gui0HLCL019}.
While this shows the feasibility of graph processing on FPGAs, these surveys only focus on the functional aspects and HPC applications.
For our survey this literature has to be reinterpreted for the database perspective.
GPU acceleration 
(\eg \cite{journals/csur/ShiZZJHLH18} for graph) is out of scope of this work.

In summary, the related surveys support the relevance and timeliness of the topic but
there is no survey on FPGA-accelerated NRDS.
The objective of this survey is to fill this gap in the intersection of FPGA acceleration and NRDS.
Their inherent differences and commonalities regarding FPGA-accelerated NRDS result in overarching design decisions and patterns that will be instrumental to current and new systems.
\subsection{Research method and contributions} 
\label{sub:methodology}
To fill the FPGA-accelerated NRDS gap in the survey literature we rely on the design science methodology \cite{DBLP:books/sp/Wieringa14} as a method to collect, summarize, and evaluate FPGA-accelerated NRDS solutions.
Our fundamental theory and motivation is: Non-relational database systems do not leverage modern FPGA hardware, from which we derive hypothesis (H1), \ie \emph{existing non-relational database systems do not realize the potential of FPGA acceleration}.
This hypothesis is tested based on an introductory system review in \cref{sec:background} which aims at analyzing existing NRDS regarding system aspects (\eg operators or scalability)
We define system aspects as abstract requirements for a system to be regarded as an NRDS.
Further, we believe that (H1) is grounded on missing fundamental research on FPGA-accelerated NRDS (cf. \cref{tab:implementations}), which leads us to hypothesis (H2), \ie \emph{there are significant gaps in current research on non-relational FPGA acceleration}.
We test this hypothesis on another observation artifact, \ie a comprehensive literature analysis in \cref{sec:literature}.
The literature analysis describes which system aspects have been previously studied and which solutions are provided along the system aspects.
In hypothesis (H3), we argue that \emph{there are patterns guiding the design of an FPGA-accelerated non-relational database system}.
To address the detected gaps and missing FPGA-accelerated NRDS we propose new practical pattern categories on how to build FPGA accelerated NRDS or accelerate existing ones.
To summarize, this survey makes the following contributions:
\begin{itemize}
    \itemsep0em
    \item We identify \emph{system aspects} resulting from challenges in accelerator design for FPGAs and a review of existing NRDS (\cref{sec:background}).
    \item We provide short solution summaries (\emph{implementations}), a table classifying references by system aspect, and a list of \emph{gaps} resulting from a comprehensive literature search (\cref{sec:literature}).
    \item We extract patterns from the literature with regard to \emph{tasks}, \emph{placements}, non-trivial accelerator \emph{design decisions}, and accelerated architecture \emph{justification} as guidelines to system architects (\cref{sec:practitioners_guide}).
    \item From the gaps in the literature we derive \emph{open research challenges} in the field of FPGA-accelerated NRDS (\cref{sec:challenges}).
\end{itemize}
\Cref{sec:summary} concludes the survey.

\subsection{Key insights -- what will the reader learn?}
In the course of this work we gain the following five key insights (highlighted as framed theorems in \cref{sec:practitioners_guide}):
\begin{enumerate}
    \itemsep0em
    \item There are three accelerator task categories (operator, data access, and communication layer acceleration) that FPGAs are currently well suited for.
    \item There are four fundamental patterns of FPGA placement (offload, SmartNIC, near-data, and socket).
    \item The accelerator task in combination with the characteristics of operators of the NRDS class are sufficient to define the FPGA placement.
    \item There are three operator switching strategies and six memory access optimization patterns guiding the development of accelerators for NRDS.
    \item A portable, relevant benchmark suite that covers all necessary artifacts is missing for robust justification of accelerator usage decisions.
\end{enumerate}
Additionally, the reader will learn about the considered NRDS classes and how they make use of accelerators (\eg scalable key-value solutions were found, but no complete database solution for others).

\section{Background}
\label{sec:background}

This section introduces fundamental concepts of FPGA hardware (\cref{sec:fpga}) and NRDS classes (\cref{sec:nosql}), required to understand the remainder of this work.
The NRDS classes are discussed in the context of well-known, commercial NRDS which we review to give an answer to hypothesis (H1) \ie \emph{existing non-relational database systems do not realize the potential of FPGA acceleration}.
Additionally, we collect important FPGA and NRDS system aspects of all NRDS classes 
in a taxonomy (\cref{fig:combined}) to guide the subsequent literature analysis.


\subsection{Field Programmable Gate Arrays}
\label{sec:fpga}

FPGAs are a processor architecture platform that map to custom architecture designs, meaning a set of logic gates and their connection (circuit design).
This is similar to application specific integrated circuits (ASICs) that are custom-made to represent a circuit design, but FPGAs are reprogrammable such that the design can be changed by a programmer at configuration time.
This reconfigurability comes at a price: from an efficiency point of view, it is always preferable to implement a design as a custom ASIC.
However, economic reasons as well as the need to adapt to changes in application behavior often prevent using ASICs.
In recent years, FPGAs emerged as accelerators for data processing (\eg \cite{series/synthesis/2013Teubner}).
They provide unique opportunities to implement functionality, like custom single instruction multiple data (SIMD) units or processing and data structure hybrids, like systolic arrays.

\begin{figure}[bt]
	\centering
	\includegraphics[width=0.75\linewidth]{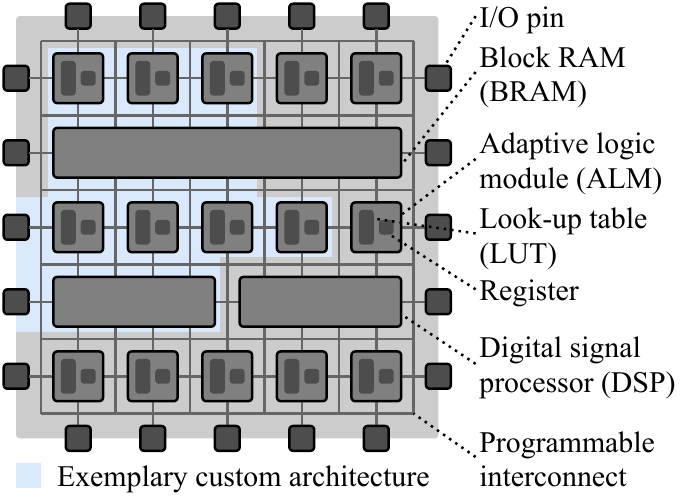}
	\caption{FPGA on-chip resources}
	\label{fig:fpga}
\end{figure}

\Cref{fig:fpga} shows an abstraction of an FPGA chip hiding vendor specifics.
On an FPGA chip, a custom architecture is mapped to a grid of resources (\eg look-up tables (LUTs) or registers) connected with a programmable interconnection network.
Each custom architecture uses a certain amount of resources upon configuration (cf. \cref{fig:fpga}), such that even multiple custom architectures might be deployed on the same chip.
Additionally, FPGAs support partial dynamic reconfiguration where only parts of the FPGA can be reprogrammed.
Data enters and leaves the FPGA chip through I/O pins.
The logic itself is implemented as adaptive logic modules (ALMs) containing a LUT with multiple inbound and one outbound bit port and a register for optional storage of the outbound bit.
Each individual LUT is programmed upon configuration.
The FPGA further contains on-chip block RAM (BRAM) in the form of SRAM memory components for fast storage of data structures.
BRAM in total is about as large as the caches on a CPU combined but finely configurable to the mapped architecture.
Lastly, the FPGA contains hardened digital signal processors (DSPs) that allow fast arithmetics on floating-point numbers.


Similar to the CPU being placed on a mainboard, FPGA chips are soldered onto a board connecting components like network ports, DRAM chips, and PCIe connectors to the FPGA.
Concerning DRAM, FPGAs support DDR3 and DDR4 as well as stacked memory enabling high bandwidth on-board data processing, like Hybrid Memory Cube (HMC).
As an accelerator, boards containing an FPGA are most often placed in existing hardware systems with a CPU -- the host system.
Traditionally, accelerators are connected with the CPU over PCIe but recently there are also cache coherent interfaces (\eg UPI, CXL, OpenCAPI).
This allows deeper integration of the FPGA into the CPU's data management and sets FPGAs apart from other accelerators like GPUs.
As an example, the Contutto system explores deeper integration of FPGAs as programmable memory buffers into the datapath between CPU and memory \cite{conf/micro/SukhwaniRHKMDSL17}.

Subsequently, the most relevant system aspects of FPGA accelerator design are discussed which are based on differences to designing CPU applications and also found to be challenging design decisions in related surveys \cite{journals/dbsk/BecherGBDGMPSTW18, journals/corr/abs-1903-06697, journals/vldb/FangMHLH20}, namely
\begin{inparaenum}[\it (i)] 
    \itemsep0em
    \item software vs. hardware design as \emph{design paradigm} (\eg \cite{journals/vldb/FangMHLH20}),
    \item tight \emph{CPU-FPGA collaboration} between existing CPU-based systems and FPGA accelerators (\eg \cite{journals/vldb/FangMHLH20}),
    \item custom \emph{memory access} controllers (\eg \cite{journals/corr/abs-1903-06697}), and
    \item comparability through a \emph{performance model} (\eg \cite{journals/dbsk/BecherGBDGMPSTW18}).
\end{inparaenum}
%


\paragraph{Design paradigm}
In principle, all LUTs ($>1,000,000$ in modern FPGAs) can operate in parallel.
This opens up a vast design space (when using hardware description languages like VHDL) compared to the well-formed design space of instruction-based processors.
While there have been efforts to simplify development with higher-level languages (\eg OpenCL), they do not seem to satisfy performance requirements for complex applications yet \cite{conf/hpec/YangSPSSH17}.
One key question of NRDS acceleration is thus, how to use constraints inherent to the NRDS class to reduce the accelerator design space without performance degradation.


\paragraph{CPU-FPGA collaboration} 
Adding an FPGA to a data processing system is justified with sufficient improvement in overall performance or cost saving.
However, most systems will still require a CPU, introducing data movement overhead between the two processors.
Thus, effective accelerator integration requires not only high utilization of the FPGA but also little data movement.
CPU-FPGA collaboration has to solve problems of task orchestration and data management.

\paragraph{Memory access} 
As one unique feature when compared to CPUs, FPGAs do not access their connected on-board memory through a deep cache hierarchy that assumes temporal and spatial locality in memory accesses.
Thus, FPGAs can implement unique caching strategies and placement of critical data on-chip.

\paragraph{Performance model}
The circuit-based programs of FPGAs have very different performance implications than instruction-based programs of CPUs.
For big circuit designs it is not easy to comprehend the amount of parallelism and make performance predictions.
Thus, performance models are about using the properties of the NRDS class to aid understanding of design decisions and comparing different approaches.

\subsection{Non-relational database systems (NRDS)}
\label{sec:nosql}
NRDS are defined according to their data models (as established in \cref{sec:introduction}).
We found five surveys on non-accelerated NRDS (cf. \cref{fig:surveys}) \cite{journals/sigmod/Cattell10, journals/csur/DavoudianCL18, GajendranS12, journals/ife/GessertWFR17, journals/asei/HeC15, conf/pervasive/HanELD11}.
According to these surveys, the predominant NRDS classes are graph, document, key-value, and wide-column.
Other NRDS classes that we do not further consider in this work are less popular ones like spatial, object-oriented or timeseries database systems.

To review well-known commercial NRDS, we selected the systems of each class that were at least mentioned three times in the NRDS surveys and combined similar systems (\eg Riak KV is similar to Amazon DynamoDB and Google Bigtable to Apache HBase).
Additionally, we added OrientDB (marked as ``expert'' in \cref{tab:nosql-systems}) as the only NRDS that successfully combines graph, document, and object-oriented database concepts despite it only being named in \cite{journals/csur/DavoudianCL18}.
\Cref{tab:nosql-systems} shows the selection of NRDS by class.
The review will not only help to briefly recall the systems' design choices and challenges but also discuss similarities.
We considered public documentation and publications if available.
Subsequently, the concepts of NRDS classes are introduced for the systems in \cref{tab:nosql-systems}. 
\begin{table}[tb]
\centering
\scriptsize
\begin{tabular}{l l r c}
    Class & System & Refs & FPGA \\
    \hline
    \hline
    \multirow{2}{*}{Graph} & Neo4j & \cite{journals/sigmod/Cattell10, journals/csur/DavoudianCL18, journals/ife/GessertWFR17, journals/asei/HeC15} & (\faThumbsOUp) \\
    & OrientDB        & \cite{journals/csur/DavoudianCL18} (expert) & \faThumbsDown \\
    \hline
    \multirow{2}{*}{Document} & CouchDB & \cite{journals/sigmod/Cattell10, journals/csur/DavoudianCL18, GajendranS12, journals/ife/GessertWFR17, journals/asei/HeC15, conf/pervasive/HanELD11} & \faThumbsDown \\
    & MongoDB & \cite{journals/sigmod/Cattell10, journals/csur/DavoudianCL18, GajendranS12, journals/ife/GessertWFR17, journals/asei/HeC15, conf/pervasive/HanELD11} & \faThumbsDown \\
    \hline
    \multirow{2}{*}{Key-value} & Redis & \cite{journals/sigmod/Cattell10, journals/csur/DavoudianCL18, journals/ife/GessertWFR17, journals/asei/HeC15, conf/pervasive/HanELD11} & \faThumbsOUp \\
    & DynamoDB & \cite{journals/sigmod/Cattell10, journals/csur/DavoudianCL18, GajendranS12, journals/ife/GessertWFR17, journals/asei/HeC15} & \faThumbsDown \\
    \hline
    \multirow{2}{*}{Wide-column} & Cassandra & \cite{journals/sigmod/Cattell10, journals/csur/DavoudianCL18, GajendranS12, journals/ife/GessertWFR17, journals/asei/HeC15, conf/pervasive/HanELD11} & (\faThumbsOUp) \\
    & HBase    & \cite{journals/sigmod/Cattell10, journals/csur/DavoudianCL18, GajendranS12, journals/ife/GessertWFR17, journals/asei/HeC15, conf/pervasive/HanELD11} & \faThumbsDown \\
\end{tabular}
\medskip
\begin{tablenotes}[flushleft]
    \centering
    \item \faThumbsOUp: productive FPGA usage, (\faThumbsOUp): explored FPGAs but no productive usage, \faThumbsDown: no FPGA acceleration mentioned
\end{tablenotes}
\caption{Commercial NRDS found in surveys}
\label{tab:nosql-systems}
\end{table}


\subsubsection{Graph}
Graph databases store entities (vertices $V$) and their relations (edges $E$).
Naively, a graph is stored as adjacency matrix $M$ of size $|V| \times |V|$ where a $1$ at position $M_{v,w}$ represents an edge between vertex $v$ and $w$.
However, currently one of the most popular representations in the literature is compressed sparse row (CSR), as a compression technique for such sparse matrices.
Moreover, graph partitioning is often applied in the literature, \eg by simple horizontal (intervals of source vertices) or vertical (intervals of destination vertices) partitioning or interval-shard \cite{conf/usenix/ZhuHC15, conf/hpec/YoungRHF16} where the graph is partitioned into $p$ equally sized intervals $I_0, I_1, ..., I_{p-1}$ of vertices and shard $S_{i,j}$ is the set of all edges between vertices in $I_i$ and $I_j$.
We look at the graph database systems Neo4j and OrientDB (cf. \cref{tab:nosql-systems}). 
Neither of those NRDS deploys FPGA acceleration in production.
However, Neo4j experimented with the Flash-accelerator CAPI SNAP \cite{capiSnap}, where an FPGA is inserted into the datapath between CPU and Flash storage to accelerate data accesses.

Neo4j\footnote{Neo4j documentation, visited 7/2020: \url{https://neo4j.com/docs/}} stores data as a property graph, where vertices and edges have properties (cf. \cref{fig:datamodel}).
It supports Cypher \cite{conf/sigmod/FrancisGGLLMPRS18} as a graph query language with a cost-based query optimizer, traversal patterns like breadth-first search (BFS), and algorithms to solve common graph problems, like shortest paths and centrality (\eg Page\-Rank).
More details on graph algorithms can be found in \cite{Even11}.
For solving graph problems, Neo4j transforms the property graph into an in-memory projection that is optimized for traversal.
Additionally, Neo4j supports different indexes.
To provide horizontal scalability, Neo4j can run as a cluster of core nodes that replicate all changes between themselves.
For additional read scaling, read replicas can be added to the cluster by registering at a core node and Neo4j drivers provide routing and load balancing.
Changes to graphs follow causal consistency, where replication to a majority of core nodes has to finish to confirm a transaction.
Optionally, ACID transaction guarantees can be enforced.
For multi-tenant usage (multiple users working on same system), Neo4j provides role-based access control (RBAC) and intra-cluster encryption, and encrypted backups provide further security.

Another well-known graph database is OrientDB\footnote{OrientDB manual, visited 7/2020: \url{https://orientdb.com/docs/last/index.html}} that allows for queries and traversals on a property graph with native support for documents.
OrientDB provides SQL-like language support with a graph extension and operators like BFS.
For scalability, data can be replicated over a cluster and availability is guaranteed by multi-master replication (similar to Neo4j) and auto discovery of nodes.
When multiple users are working on the database (OrientDB also supports RBAC), an optimistic multi-version concurrency control (MVCC) is used and ACID transaction guarantees can be applied.
OrientDB supports encryption on disk.



\subsubsection{Document}
\label{sec:document}
Document systems store formatted text documents in a document hierarchy.
The two most popular document formats are Extensible Markup Language (XML) \cite{BrayPSMY00} and JavaScript Object Notation (JSON) \cite{Bray14}.
All documents in a document store adhere to such a fixed formatting and are parsed upon ingestion from a string representation for quick processing afterwards.
For XML stores, there is the XML Path Language (XPath) \cite{ClarkD99} as a query language. 
Apache CouchDB and MongoDB 
are the most-referenced document NRDS (cf. \cref{tab:nosql-systems}).
We did not find FPGA acceleration options for any of these two document stores.

Apache CouchDB\footnote{Apache CouchDB documentation, visited 7/2020: \url{https://docs.couchdb.org/en/stable}} is a JSON document store operating as a cluster of masters with bidirectional, asynchronous replication.
The API allows create, read, update, and delete (CRUD) operators on documents and more advanced view models for filtering and aggregation.
Queries can be accelerated with B-tree indices.
Writes to the database are isolated with MVCC and ACID transaction guarantees can be enforced.
Like the graph databases, Apache CouchDB supports RBAC and provides security features, like update validation.

MongoDB\footnote{The MongoDB manual, visited 7/2020: \url{https://docs.mongodb.com/manual}} is another JSON document store with similar to Apache CouchDB query possibilities but also adds capabilities for fulltext search and spatial queries.
Availability is provided by a master-slave cluster setup, where nodes are placed into a replication set of a master that handles all writes.
Writes are atomic, and recently multi-document transactions were added to the system.
If the master fails, a new one is elected among the replicas.
Data can also be sharded over multiple replication sets for scalability.
Similar to the other systems, MongoDB provides RBAC for authentication of users.
For security, communication between users and the database can be SSL-encrypted.

\subsubsection{Key-value}
Key-value stores operate on pairs of a key used for quick lookup and a value of arbitrary data.
They may be used as an underlying persistence layer of another database or standalone depending on the use case.
We subsequently discuss Redis and Amazon DynamoDB (cf. \cref{tab:nosql-systems}).
While Amazon DynamoDB has no accelerator options, there is a recent extension for Redis by Algo-Logic \cite{redisFpga} with the FPGAs directly attached to the network for increased throughput, lower latency, and less energy consumption.

Redis\footnote{Redis documentation, visited 7/2020: \url{https://docs.redislabs.com/latest/index.html}} may be used as an in-memory or a persistent key-value store, with a typical operator set with \texttt{set} and \texttt{get} operators.
Key-value pairs are inserted into a hash table, where the key hash is used as the index in the table for fast lookup of values, but complex queries are not supported.
Similar to the graph and document stores, Redis uses a master-slave setup with sharding for scalability.
Additionally, it supports request routing on a proxy node and optional waiting for replication for consistency.
Redis has security features like SSL encryption and action auditing.

Amazon DynamoDB\footnote{Amazon DynamoDB developer guide, visited 7/2020: \url{https://docs.aws.amazon.com/amazondynamodb/latest/developerguide/Introduction.html}} \cite{conf/sosp/DeCandiaHJKLPSVV07} features a distribution scheme without a master.
The key hashes denote a circular space and each node is randomly placed in this space upon entrance to the cluster. 
Each node is responsible for the keys preceding it in counter-clockwise order in this space and answers all requests to that partition.
Data is replicated in clockwise order to a fixed number of nodes and upon failure, the node after a failed one is responsible for the failed nodes partition of the circular space.
For load balancing, Amazon DynamoDB places many more virtual nodes on the circular space than there are nodes in the cluster.
Multiple virtual nodes are then handled by each physical node.
Amazon DynamoDB supports eventual consistency, MVCC, and is integrated with the RBAC of AWS accounts.
For security, it features encryption and user authentication.



\subsubsection{Wide-column}
Wide-column databases also store data as pairs of key and value.
However, different to key-value databases, the key has two predefined parts: a row name and a column name.
Wide-column databases present their data as tables to the user but unlike relational databases the tables are not materialized, unstructured, and every row may have arbitrary columns.
The most-referenced wide-column systems are Apache Cassandra and Apache HBase (cf. \cref{tab:nosql-systems}).
For Apache Cassandra, FPGAs may be used to accelerate the data accesses where the FPGA denotes a data proxy \cite{cassandraFpga}.



Apache Cassandra\footnote{Apache Cassandra Documentation, visited 7/2020: \url{https://cassandra.apache.org/doc/latest/}} \cite{journals/sigops/LakshmanM10} uses a multi-dimensional map, indexed with a key (everything with the same key constitutes a row) and columns that are grouped into column families.
It supports \texttt{insert}, \texttt{get}, and \texttt{delete} operations and additionally has a Cassandra query language (CQL) comparable to SQL.
Similar to Amazon DynamoDB, the key hashes are used as a circular space to partition the data to the nodes in the cluster but load balancing is done by moving nodes in the circular space when imbalance is detected.
Consistency is guaranteed with a replication quorum and RBAC is supported as well as SSL-encryption.

Apache HBase\footnote{Apache HBase reference guide, 7/2020: \url{https://hbase.apache.org/book.html}} is a wide-column store based on Apache HDFS.
The supported operator set as well as the binary representation is similar to Apache Cassandra, but Apache HBase distributes updates in a master-slave fashion.
Similar to all other systems we looked at, Apache HBase supports RBAC.







\subsection{Summary -- NRDS on reprogrammable hardware}
\label{sec:nosql_arch}
\begin{figure*}[bt]
	\centering
	\includegraphics[width=0.8\textwidth]{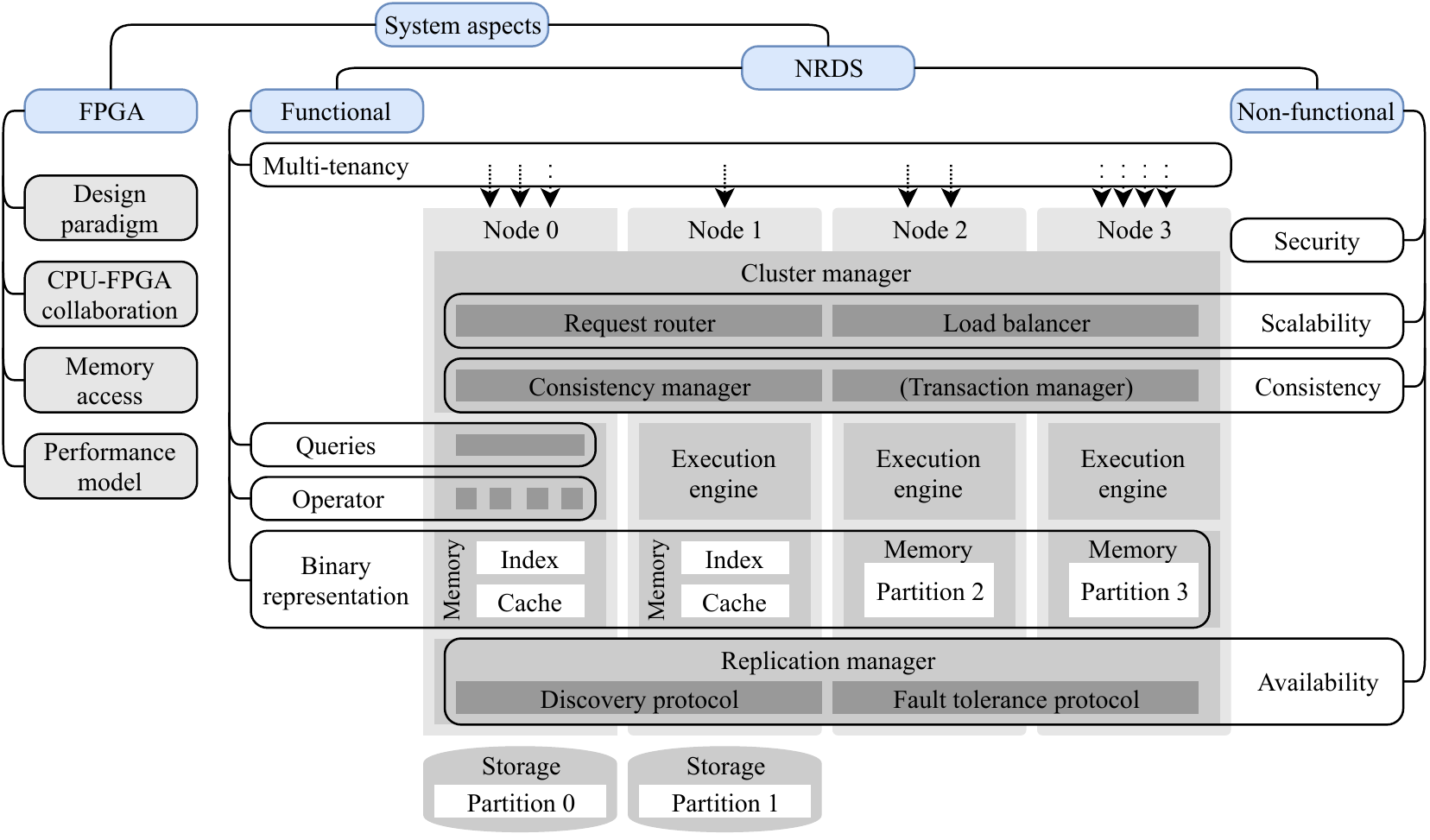}
	\caption{Taxonomy of system aspects along common NRDS architecture}
	\label{fig:combined}
\end{figure*}

Although the architectures of the systems reviewed exhibit different feature sets, they cover a shared set of system aspects.
For example, Neo4j provides an execution engine with Cypher query capabilities while Apache Cassandra provides them with CQL.
Scalability in Amazon DynamoDB and Apache Cassandra is achieved with partitioning data with a circular hash space while \eg MongoDB and Redis distribute work in a master-slave fashion.
While the query languages and scalability schemes might differ in their concrete implementation, queries and scalability are two integral system aspects of any NRDS.

\Cref{fig:combined} shows a generalized shared architecture as a component view of the reviewed NRDS capturing the system aspects we found in the system review.
The architecture is based on a set of nodes in a networked environment, forming a cluster.
Multiple different users or tenants send requests to the system, which distributes work with the request router and load balancer in the cluster manager.
Changes to the underlying data are kept consistent by the consistency manager, which performs concurrency control and atomic writing.
Optionally the transaction manager provides transactions (\eg ACID) touching multiple data elements.
The execution engine finally processes queries -- made up of operators -- on the data stored in the system. 
Queries performance may be improved with indexes.
Persistent storage is optional, but data is partitioned over the nodes.
Lastly, the replication manager aids discovery of new nodes in the system and fault handling.
While the components in general might be similar to scale-out relational database systems, NRDS lay different emphasis on components because of different application requirements (cf. \cref{sec:introduction}).

As an overlay in \cref{fig:combined}, we show the FPGA system aspects from \cref{sec:fpga} and functional and non-functional system aspects defining NRDS.
A complete system would have to satisfy all (or most of the) combined system aspects.
FPGAs are not well established enough in NRDS to place them in the shared architecture, such that their system aspects are added separately.
The aspect taxonomy will be used in the literature analysis (\cref{sec:literature}) to classify literature.
In summary, there are some first experiments with FPGAs in NRDS \cite{capiSnap, redisFpga, cassandraFpga} that show the potential and feasibility, but the majority of systems has not considered FPGAs yet.
Thus, we confirm hypothesis (H1) \ie \emph{existing non-relational database systems do not realize the potential of FPGA acceleration}.

\section{Literature Review}
\label{sec:literature}

In this section we conduct a literature analysis in order to answer hypothesis (H2), \ie \emph{there are significant gaps in current research on non-relational FPGA acceleration}, as set out in \cref{sub:methodology}.
The hypothesis raises three questions to be investigated in the literature analysis: (a) What are the most relevant of the identified NRDS classes? 
(b) Are there any system aspects that are not yet covered by literature? (c) Do existing approaches provide solutions to these topics?

The literature analysis is based on the guidelines described in \cite{kitchenham2004procedures}.
The primary selection of references is conducted in the domain of each individual NRDS class and with a focus on research articles (no patents and citations).
This results in $351$ hits before the following selection criteria are applied:
\begin{inparaenum}[\it (i)] 
    \itemsep0em
    \item relation to computer science, FPGA, reconfigurable hardware, and acceleration,
    \item focus on data processing (excluding for example robotics, image processing, or graph-based FPGA design),
    \item availability of the document,
    \item written in English,
    \item published (excluding Master and PhD theses). 
\end{inparaenum}
Overall, this results in $89$ selected articles relevant to this survey. 

Notably, we did not find dedicated literature for FPGA-acceleration of wide-column which, however, can be seen as very similar to key-value (cf. \cite{journals/csur/DavoudianCL18}). 



\subsection{Processing of selected literature -- NRDS classes and trends}
All selected articles 
were categorized by NRDS class giving an answer to question (a) What are the most relevant of the identified NRDS classes?

\begin{figure}[bt]
	\centering
	\includegraphics[width=\linewidth]{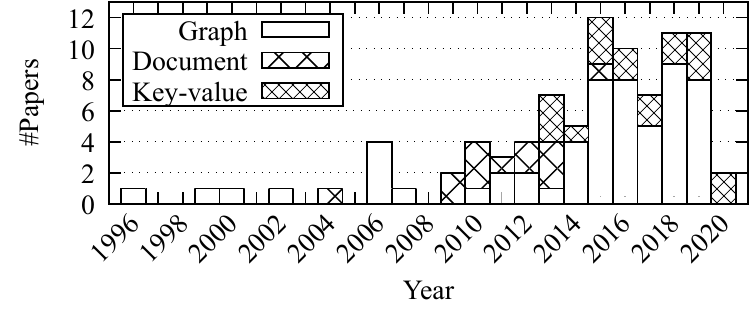}
	\caption{Topics in literature over time}
	\label{fig:histogram}
\end{figure}
\Cref{fig:histogram} depicts the distribution of NRDS classes over time.
It can be seen that hardware-accelerated graph processing already plays an early role from $1999$--$2002$.
After three years with only one article for document, graph sparks in the years $2006$ and $2007$.
From $2009$--$2013$ document moved into the focus, with graph joining in again from $2010$ onward.
Recently (\ie $2013$--$2019$), the picture seems to change, turning more or less exclusively to graph and key-value.

From \cref{fig:histogram} it can be concluded that most of the non-relational hardware work occurred in literature after $2006$ with a dominance of graph.
Apparently document lost significance, and other classes like key-value gained more attention.
From the dominance of graph and the very different key-value type we conclude that a more fine-grained analysis of the mentioned database classes is meaningful.



\subsection{Literature summaries}
This section summarizes the approaches identified in the literature search, thus addressing questions (b) Are there any system aspects that are not yet covered by literature? and (c) Do existing approaches provide solutions to these topics?
We organize the summaries chronologically (cf. \cref{fig:histogram}) and by NRDS class.
Additionally, we structure the literature by their strongest contribution(s) to the system aspects (cf. \cref{fig:combined}).
\Cref{tab:implementations} summarizes the found solutions and subclassification we identified for each NRDS class and system aspect pairing which we briefly introduce subsequently.

%
\begin{table*}[ht]
\centering
\scriptsize
\begin{tabular}{l p{4cm} p{4cm} p{4cm}} 
 & \textbf{Graph} & \textbf{Document} & \textbf{Key-value}\\ 
 \hline
 \hline
 \textbf{NRDS -- functional} & & & \\
 \hline
 Operator
 &
 \textbf{SP} \cite{DBLP:conf/fpl/BetkaouiWTL12, conf/asap/BetkaouiWTL12, conf/ipps/BondhugulaDFWS06, journals/iet-cdt/JagadeeshSL11, journals/tcas/LeiDLX16, journals/tjs/MilovanovicMBT07, conf/pdpta/TakeiHK14, TakeiHK15, conf/ipps/ZhouCP15};
 \textbf{BFS} \cite{DBLP:conf/dac/FinnertySLL19, journals/iracst/LeiRG15, DBLP:conf/fpl/UmurogluMJ15, DBLP:conf/fpt/WangJXP10};
 \textbf{MM} \cite{DBLP:conf/fpga/BestaFBLH19}; 
 \textbf{PR} \cite{DBLP:conf/reconfig/ZhouCP15}; 
 \textbf{Cent.} \cite{conf/fpl/GiefersSP16}
 & 
 \textbf{XML Parser} \cite{conf/fpga/DaiNZ10, HuangJWWP14, conf/fpt/Sidhu13};
 \textbf{Filter} \cite{conf/ispass/ChalamalasettiMVWR12};
 \textbf{XPath} \cite{conf/reconfig/El-HassanI10, journals/corr/abs-0909-1781, conf/hipeac/MoussalliSNT10};
 \textbf{Twig} \cite{conf/icde/MoussalliSNT11}
 &
 \textbf{Insert} \cite{conf/fpl/LiangYKW16};
 \textbf{Hash} \cite{conf/hpcc/LiuZJJ19}
 \\
 \hline
 Binary representation	& 
 \textbf{HP} \cite{conf/fccm/ZhouCP16};
 \textbf{VP} \cite{conf/fpl/ChenBCHHWC19};
 \textbf{IS} \cite{conf/fpga/DaiCWY16};
 \textbf{CSR} \cite{conf/ipps/AttiaJTJZ14}; 
 \textbf{HWM} \cite{conf/fpga/Huelsbergen00, conf/fpl/MencerHH02}; 
 \textbf{Other} \cite{SklyarovSP06, conf/bigdataconf/TariqS18, conf/hpcc/WangHZZPX15, journals/jsa/WangZH16, conf/icws/XuWGJLZ18} &  & 
 \textbf{Hash table} \cite{conf/hotcloud/BlottKLVBI13, conf/fpl/IstvanABV13, conf/ipps/TongZP15, YangKSKP20}; 
 \textbf{CAM} \cite{conf/hoti/LockwoodM15} \\
 \hline
 Queries	& \textbf{Instructable processor} \cite{conf/asap/Kapre15} & 
 \textbf{Skel. autom.} \cite{conf/sigmod/TeubnerWN12}; 
 \textbf{Other} \cite{conf/www/LunterenEBCL04} & n/a \\
 \hline
 Multi-tenancy  &  &  & \textbf{Token bucket} \cite{conf/fpl/IstvanAS18} \\
 \hline
 \hline
 \textbf{NRDS -- non-functional} & & & \\
 \hline
 Scalability	& \textbf{Part.} \cite{DBLP:conf/reconfig/AttiaGTJZ15, conf/other/BabbFA96, conf/fpga/DaiHCXWY17} &  & 
 \textbf{Repl.} \cite{journals/pvldb/IstvanSA17};
 \textbf{Part.} \cite{QiuXLYLWYCGLS20, journals/pvldb/XuLJLHA16} \\
 \hline
 Availability	&  &  & 
 \textbf{Repl.} \cite{journals/pvldb/IstvanSA17} \\
 \hline
 Consistency	& 
 \textbf{Locking} \cite{conf/fpga/MaZC17} &  & 
 \textbf{MVCC} \cite{conf/fpl/RenLSXQLYWYCH19} \\
 \hline
 Security   	&  &  & \\
 \hline
 \hline
 \textbf{FPGA} & & & \\
 \hline
 Design paradigm	& 
 \textbf{VC} \cite{DBLP:conf/fpl/EngelhardtS16, DBLP:conf/fccm/NurvitadhiWWHNH14, DBLP:conf/isca/OzdalYKAGBO16, WeiszNH13, DBLP:journals/corr/abs-1806-00751}; 
 \textbf{EC} \cite{journals/tpds/ZhouKPSW19, conf/cf/ZhouKZP18}; 
 \textbf{Hybrid} \cite{journals/corr/abs-1806-11509}; 
 \textbf{BSP} \cite{journals/tcad/AyupovYOKBO18, conf/fccm/DeLorimierKMRERUKD06};
 \textbf{Other} \cite{conf/other/BabbFA96, DBLP:conf/ipps/DandalisMP99, DBLP:conf/fpga/OguntebiO16} &  & n/a \\
 \hline
 CPU-FPGA collaboration	& \textbf{Socket} \cite{conf/fccm/BondhugulaDDFWSS06, DBLP:conf/fpl/UmurogluMJ15, DBLP:conf/fccm/WangHN19, conf/sbac-pad/ZhouP17}; \textbf{Near-data} \cite{DBLP:journals/pvldb/LeeKYCHNNJ17} & \textbf{Socket} \cite{conf/fpl/VanderbauwhedeAM09}; \textbf{Near-data} \cite{journals/tods/TeubnerWN13}; \textbf{PCIe} \cite{journals/sigarch/VanderbauwhedeFCM13} & \textbf{Near-data} \cite{journals/cal/LavasaniAC14, conf/icfpt/XieQYW19}; \textbf{DMA} \cite{conf/sosp/LiRXLXPCZ17, conf/fpt/QiuLXYW18} \\
 \hline
 Memory access	& 
 \textbf{Req. merging} \cite{conf/fpga/KhoramZSL18}; 
 \textbf{Caching} \cite{DBLP:conf/fpga/ShaoLHL019, DBLP:conf/fpga/ZhangKL17}; 
 \textbf{Data placem.} \cite{DBLP:conf/fpga/ZhangL18};
 \textbf{Other} \cite{DBLP:conf/fpt/BetkaouiTLP11, DBLP:journals/ieiceee/NiDZLW14, DBLP:conf/islped/YanHLALMDYZF019} &  & 
 \textbf{Bloom filter} \cite{conf/vlsi-dat/ChoC14}; 
 \textbf{Flash} \cite{conf/hotstorage/BlottLKV15, journals/pvldb/XuLJLHA16} \\
 \hline
 Performance model & \textbf{Parallelism} \cite{conf/ipps/BondhugulaDFWS06, DBLP:conf/fpga/ShaoLHL019}; \textbf{Bottleneck} \cite{conf/fpga/DaiCWY16, conf/fpga/DaiHCXWY17, DBLP:conf/fpl/EngelhardtHS18, DBLP:conf/fpga/ZhangKL17} &  & \textbf{Bottleneck} \cite{QiuXLYLWYCGLS20} \\
\end{tabular}
\medskip
\begin{tablenotes}
    \centering
    \item[*] SP: Shortest path, BFS: Breadth-first search, MM: Maximum matchings, PR: RageRank, Cent.: Centrality, HP: Horizontal partitioning; VP: Vertical partitioning, IS: Interval-shard, CSR: Compressed sparse row, HWM: Hardware mapping, Part.: Partitioning, VC: Vertex-centric, EC: Edge-centric, BSP: Bulk-synchronous parallel, Req. merging.: Request merging, Data placem.: Data placement; Skel. autom.: Skeleton automaton; CAM: Content addressable memory, Repl.: Replication, MVCC: Multi-version concurrency control, DMA: Direct memory access
\end{tablenotes}
\caption{Contributions by domain and system aspect}
\label{tab:implementations}
\end{table*}


\subsubsection{Graph}

According to the timeline (cf. \cref{fig:histogram}), the first non-relational accelerator solutions were provided for graph representing mostly HPC solutions not specifically tuned towards NRDS.

\paragraph{Operator} In the literature, we identified solutions for five different operators which we discuss in the following order: shortest path, breadth-first search, maximum matchings, page rank, and centrality.

\labeltitle{Shortest path}
Bondhugula et al. present a tiled Floyd-Warshall implementation solving the all-pairs shortest-paths problem using a pipeline of $B$ processing elements (PE) which can each process $l$ elements of the adjacency matrix \cite{conf/ipps/BondhugulaDFWS06}.
They build a performance model with these two parameters where $B$ is constrained by FPGA resources and $l$ is constrained by I/O bandwidth.

Jagadeesh et al. use a parallel, synchronous Bellman-Ford algorithm where each PE on the FPGA represents one node in the graph (severely limiting graph size) \cite{journals/iet-cdt/JagadeeshSL11}.
Additionally, they model the internal computation time of their shortest path computation unit in cycles for performance prediction.
%
A parallel implementation of Bellman-Ford that considers conflicting vertex updates 
is presented by Zhou et al. \cite{conf/ipps/ZhouCP15}.
The edges are processed in parallel and conflicts are resolved by caching updates on-chip until they are applied in main memory.
To optimize for sequential reads, the edges are sorted by destination vertex.

In \cite{conf/pdpta/TakeiHK14, TakeiHK15}, Takei et al. combine Dijkstra with a SIMD distance comparison unit.
A restriction of the approach is that all nodes have to fit into on-chip memory.
Lei et al. propose an eager Dijkstra algorithm based on their own memory overflow extension of a priority queue and three memory channels (for overflow queue, graph, and output data) \cite{journals/tcas/LeiDLX16}.
This resolves the graph size restriction of the earlier approach by Takei et al.

In \cite{journals/tjs/MilovanovicMBT07}, the authors propose a solution for all-pairs shortest-path by defining a partitioning that allows for processing the graph with a bidirectional systolic array with an optimal number $|V|$ of PEs.
Betkaoui et al. solve all-pairs shortest-paths with parallel BFS kernels \cite{DBLP:conf/fpl/BetkaouiWTL12, conf/asap/BetkaouiWTL12}.
Multiple parallel PEs issue many non-blocking memory requests to take advantage of the multi-channel memory system of their particular FPGA setup.
\labeltitle{Breadth-first search (BFS)}
Wang et al. propose a solution for a parallel BFS with message passing on a fully-connected network between interval-partitioned soft cores \cite{DBLP:conf/fpt/WangJXP10}.
The traversal levels are synchronized with a floating barrier. 
The number of random memory accesses is reduced by keeping the visited status in on-chip memory. 
Additionally, their approach allows switching traversal patterns (bottom-up and top-down) per level.
TorusBFS \cite{journals/iracst/LeiRG15} proposed by Lei et al. implements torus network-based message-passing between PEs in an interval-partitioned graph.
The PEs are similar to those in \cite{DBLP:conf/fpl/BetkaouiWTL12}, and auxiliary data structures are stored in BRAM.

In \cite{DBLP:conf/fpl/UmurogluMJ15}, BFS iterations are substituted by matrix-vector operations on a Boolean semi-ring (i.e.\ multiply and add are substituted with logical \texttt{and} and \texttt{or}).
The data is horizontally partitioned but the matrix and vectors are never materialized.
%
Dr. BFS \cite{DBLP:conf/dac/FinnertySLL19} uses vertical partitioning to fit metadata of large graphs into on-chip memory. 
The tasks of computation and data access are separated into two different modules having a data access module for every memory bank. 
The computation modules use a pipelined combination network for sequential burst operations.

\labeltitle{Maximum matchings}
Besta et al. propose a solution for maximum matchings in a graph which describes the maximum size set of edges that do not share a vertex \cite{DBLP:conf/fpga/BestaFBLH19}.
Their substream-centric approach divides the incoming stream of edges by their weight into substreams, processes them in parallel, and merges the results.

\labeltitle{PageRank}
Zhou et al. propose an implementation of PageRank (a measure for importance of vertices in a graph) split up into a scatter and gather phase using a source vertex interval (\ie horizontal) graph partitioning with vertex, edge, and update sets for each partition \cite{DBLP:conf/reconfig/ZhouCP15}.
The edge set of each partition is sorted by destination vertex to reduce the number of random writes.

\labeltitle{Centrality}
In \cite{conf/fpl/GiefersSP16}, the authors propose a stochastic matrix function estimator written in OpenCL and apply it to the subgraph centrality problem. 
Subgraph centrality (like PR) is another measure for importance of vertices in a graph.

\paragraph{Binary representation} 
We found approaches that leverage different partitioning schemes (horizontal, vertical, interval-shard), an optimization of CSR for BFS, and other binary representations like hardware mapping.

\labeltitle{Horizontal partitioning}
In \cite{conf/fccm/ZhouCP16} a horizontal partitioning scheme is proposed with an improved data layout enabling more sequential write accesses.
This is achieved by sorting the edges in each partition by destination vertex which also allows update merging.


\labeltitle{Vertical partitioning}
Chen et al. provide a solution that features a layout improvement of the partitioned data by storing edges inbound to the current vertex set and not storing an update list (i.e.\ directly streamed) \cite{conf/fpl/ChenBCHHWC19}.
The data is shuffled to graph PEs.


\labeltitle{Interval-shard}
FPGP \cite{conf/fpga/DaiCWY16} is an edge-centric graph-processing framework based on the interval-shard graph partitioning scheme.
Shards $S_{i,j}$ and their corresponding inbound $I_i$ and outbound $I_j$ vertex intervals are stored one after another in on-chip memory and processed by multiple PEs.


\labeltitle{Compressed sparse row (CSR)}
The CyGraph architecture by Attia et al. proposes new optimizations for utilizing the memory bandwidth in parallel BFS by a custom CSR representation \cite{conf/ipps/AttiaJTJZ14}.
The visitation status of vertices is encoded in the row index which is replaced by the level after visitation (in-place BFS result).


\labeltitle{Hardware mapping}
In \cite{conf/fpga/Huelsbergen00}, the complete graph with editable vertices and edges is represented as logic on the FPGA. 
An extension by Mencer et al. \cite{conf/fpl/MencerHH02} draws parallels to content-addressable memory (CAM) and extends the idea of mapping the adjacency matrix to hardware by extending it to multi-context graph processing.



\labeltitle{Other}
Skylarov et al. represent the graph as a matrix and use matrix operations to calculate graph colorings \cite{SklyarovSP06}. 
%
Wang et al. propose an edge-centric graph streaming model on an FPGA for partitioned graphs to deal with load balancing issues of skewed graphs \cite{conf/hpcc/WangHZZPX15,journals/jsa/WangZH16}. 
The $kp$ partitions are assigned to $k$ PEs in a pre-processing step such that every PE processes approximately the same number of edges.
The graph is compressed and streamed which results in a high effective bandwidth.
%
When the application permits graph sampling, the data volume and irregular memory access can be tamed with a data structure derived from CSR \cite{conf/bigdataconf/TariqS18}.
The novel data structure allows storing multiple graphs and removal of vertices and edges with a second pointer array.
Xu et al. propose a service-oriented accelerator that does asynchronous processing of BFS \cite{conf/icws/XuWGJLZ18}.
This is very different to other approaches in that batches of so called Batch Row Vectors are streamed into the accelerator and worked on.
A Batch Row Vector contains for each edge in the batch the start and destination vertex and a property for the source and destination vertex.



\paragraph{Queries}
Queries in the context of accelerator design are about flexible chaining of operators.
We did not find any accelerator designs explicitly handling query workloads.
However, there is an instructable processor design that could be extended for query processing.

\labeltitle{Instructable processor}
GraphSoC is a custom graph processor built from a 2D array of softcore processors (with send, receive, accumulate, and update instructions) connected by a packet-switched network \cite{conf/asap/Kapre15}.


\paragraph{Scalability}
There are no complete solutions for scalability either.
However, there are three articles on graph partitioning to leverage multi-FPGA setups.

\labeltitle{Partitioning}
Babb et al. presented an early work on scalability by compiling graph problems to whole arrays of FPGAs \cite{conf/other/BabbFA96}.
Virtual wires are used in between the FPGAs resulting in a multi-FPGA computing fabric.
Attita et al. propose a vertex-centric graph processing framework based on the gather-apply-scatter (GAS) principle \cite{DBLP:conf/reconfig/AttiaGTJZ15}.
With partitioning and message passing, the workload can be distributed to multiple FPGAs.
ForeGraph \cite{conf/fpga/DaiHCXWY17} is an edge-centric graph processing approach as a multi-FGPA extension to \cite{conf/fpga/DaiCWY16}.
For $p$ FPGAs, the graph is partitioned into $p$ intervals resulting in $p^2$ shards.
Each FPGA stores one interval and its outgoing $p$ shards, additionally partitions its subgraph into sub-intervals and sub-shards, and subsequently does the same processing as \cite{conf/fpga/DaiCWY16} on the sub-shards.
Updates to foreign intervals are propagated to the corresponding FPGA over the network.



\paragraph{Consistency} 
The conflict management for highly concurrent systems (transactions on graphs) can be problematic.
We only found one article proposing a locking scheme to achieve isolation.

\labeltitle{Locking}
Ma et al. define a multi-threaded graph processing engine on FPGA using a global, transactional shared memory which allows fine-granular locking with an address signature table \cite{conf/fpga/MaZC17}. 


\paragraph{Design paradigm} 
We found multiple graph accelerator design paradigms: vertex-, edge-centric, hybrid, and BSP, among others.
Vertex- and edge-centric describe orthogonal approaches of traversing graphs: while the vertex-centric approach works on outgoing edges of active vertices, the edge-centric approach loads all edges from memory and discards those not needed currently.

\labeltitle{Vertex-centric}
GraphGen \cite{DBLP:conf/fccm/NurvitadhiWWHNH14} compiles graph algorithms from a domain specific language to RTL.
The algorithms are built from user-defined instructions.
RTL implementations of these instructions have to be provided together with an update function and a description on how those instructions play together.
Weisz et al. provide programmability by using GraphGen and CoRAM in combination \cite{WeiszNH13}.
CoRAM allows to port the accelerator architecture to both Intel and Xilinx FPGAs.
%
GraVF \cite{DBLP:conf/fpl/EngelhardtS16} shows how to compile Migen definitions to hardware.
Ozda et al. provide a solution based on a configurable architecture template for vertex-centric graph algorithms \cite{DBLP:conf/isca/OzdalYKAGBO16}.
In \cite{DBLP:journals/corr/abs-1806-00751}, conflicting updates on one vertex are resolved by accumulating updates in one cycle, parallelizing conflicting vertex updates, and removing sequential application of atomic protection.

\labeltitle{Edge-centric}
HitGraph \cite{journals/tpds/ZhouKPSW19, conf/cf/ZhouKZP18} (based on \cite{DBLP:conf/reconfig/ZhouCP15, conf/fccm/ZhouCP16}) compiles edge-centric algorithms to RTL. 
The processing logic is split up into multiple PEs processing the graph by alternating scatter and gather phases. 
The vertices are partitioned and buffered in BRAM and partitions are skipped if they do not contain active vertices.


\labeltitle{Hybrid}
Chengbo proposes a hybrid approach where there is a vertex-centric and an edge-centric module on the FPGA programmed in OpenCL \cite{journals/corr/abs-1806-11509}. 
A dispatcher switches the modules depending on the workload.
The graph is vertically partitioned.

\labeltitle{Bulk-synchronous parallel (BSP)}
DeLorimier et al. propose an accelerator, mapping the graph to sparse matrix operations executed in a BSP fashion \cite{conf/fccm/DeLorimierKMRERUKD06}.
The accelerator is split up into compute leaves with BRAMs attached.
Data between the compute leaves is communicated over a network on chip.
%
In \cite{journals/tcad/AyupovYOKBO18}, a design methodology based on the BSP model is proposed.
Common architectural features are represented as templates which are specified with user-defined functions for GAS. 
All data flow is handled by the template.

\labeltitle{Other}
In \cite{conf/other/BabbFA96}, problems on directed graphs are reformulated as closed semiring problems and compiled onto multiple FPGAs.
For a graph instance the edges are mapped to summations and vertices are mapped to minimum operators.
Dandali et al. address long synthesis times by a skeleton compilation with precompiled blocks that are adapted to the problem instance \cite{DBLP:conf/ipps/DandalisMP99}.
One PE is used for a vertex and has to fit onto the FPGA.
GraphOps \cite{DBLP:conf/fpga/OguntebiO16} is a general graph dataflow library that provides graph-specific building blocks for the generation of FPGA designs.
It includes a locality-optimized property array.




\paragraph{CPU-FPGA collaboration} 
We found the following contributions running on CPU-FPGA heterogeneous platforms with different task assignments.

\labeltitle{Socket}
Bondhugula et al. propose a shortest-path hardware kernel communicating with the CPU via a shared memory region \cite{conf/fccm/BondhugulaDDFWSS06}.
Another part of the solution is a graph data layout for the CPU to reduce cache misses.
Umuroglu et al. propose an approach that distributes BFS iterations between CPU and FPGA: the iterations with few active vertices are performed on the CPU while the other iterations are performed on the FPGA \cite{DBLP:conf/fpl/UmurogluMJ15}.
Similarly, Zhou et al. address the respective drawbacks of vertex- and edge-centric graph processing by switching between the paradigms during execution \cite{conf/sbac-pad/ZhouP17}.
The graph is horizontally partitioned (cf. \cite{conf/fccm/ZhouCP16}) and the paradigm is chosen individually for each partition in each iteration based on its active vertex ratio.
Partitions with few active vertices are processed by the CPU in a vertex-centric paradigm while partitions with many active vertices are processed by the FPGA in an edge-centric paradigm.
%
Wang et al. propose a general graph processing approach with a novel worklist (priority queue) based graph computation and software scheduler (reorders vertices to be processed) \cite{DBLP:conf/fccm/WangHN19}.
The FPGA inserts work items (vertices) into a pre-scheduled queue and the CPU reorders them into a scheduled queue.

\labeltitle{Near-data}
ExtraV \cite{DBLP:journals/pvldb/LeeKYCHNNJ17} proposes graph virtualization.
The CPU accesses the data through a cache coherent FPGA attached to an SSD storing the graph. 
Transparently to the CPU, the FPGA applies compression and multi-versioning to writes and decompression and filtering to reads.


\paragraph{Memory access}
For memory access, we found solutions for request merging, custom caching, and custom data placement among others.

\labeltitle{Request merging}
A CAM-based approach for the BFS memory access problem is proposed in \cite{conf/fpga/KhoramZSL18}.
The solution is an architecture-aware software graph clustering algorithm that reduces bandwidth requirements for random requests to visited flags.
The clustering is applied as an offline pre-processing step.
A memory unit merges multiple requests (cache for storing recently checked flags implemented with CAM).


\labeltitle{Caching}
Zhang et al. propose a map-reduce based BFS approach on HMC memory \cite{DBLP:conf/fpga/ZhangKL17}.
With a performance model they find the bottleneck in scanning the bitmap of the current frontier.
Thus, a second caching layer for the bitmap is introduced where one bit in the caching layer represents the aggregate of many bits in the RAM bitmap.
%
FabGraph \cite{DBLP:conf/fpga/ShaoLHL019} is an extension of ForeGraph \cite{conf/fpga/DaiHCXWY17} by a two-level vertex caching (level L1 attached to pipe\-lines and shared L2; L1 only communicates with L2).
The vertices of the current graph partition are stored in L2 and replaced in Hilbert order such that vertices can be used for multiple graph portions.

\labeltitle{Data placement}
Zhang et al. address the problem of redundant memory accesses caused by high-degree vertices for graph traversals \cite{DBLP:conf/fpga/ZhangL18}.
They correlate vertex degree with data access frequency and propose degree-aware data placement and degree-aware adjacency list compression (Exp-Golomb variable length coding) combined with hybrid traversal approach on HMC memory.


\labeltitle{Other}
Betkaoui et al. address stalling pipelines caused by memory access latency through a memory crossbar to share off-chip memory \cite{DBLP:conf/fpt/BetkaouiTLP11}.
The solution issues many parallel memory requests and decouples memory access and execution units.
Ni et al. accelerate BFS by applying a horizontal partitioning allowing to distribute the graph and its associated metadata over multiple memory channels \cite{DBLP:journals/ieiceee/NiDZLW14}.  
In this way, multiple PEs can traverse the graph in parallel utilizing a high memory bandwidth.
Upon level synchronization, active vertices are exchanged between the PEs.
%
A memory access improvement for vertex-centric graph processing is given by Yan et al. \cite{DBLP:conf/islped/YanHLALMDYZF019}.
Random memory accesses are sequenced and graph pruning is applied to prevent ongoing traversal from the leaves.
The solutions further includes an online pre-processing step during the apply phase when bandwidth is under-utilized.

\paragraph{Performance model}
Performance models are about understanding effects of design decisions on a conceptual level. 
We found the following solutions on modeling parallelism and bottlenecks.

\labeltitle{Parallelism}
Bondhugula et al. define a performance model for parallelism in two orthogonal parameters $B$ number of PEs ($B$ loops at once), and $l$ denoting the number of operators in PE ($l$ elements at once) \cite{conf/ipps/BondhugulaDFWS06}.
Then constraints for different FPGA resources are modelled for the two parameters. 
Shao et al. model the runtime of their design as the sum of the time of the vertex transmission and the time of the edge streaming \cite{DBLP:conf/fpga/ShaoLHL019}.
Both are modeled dependent on the internal parallelism parameters of their design.
The model is used to determine the size of L1 and L2 caches in the design.



\labeltitle{Bottleneck}
FPGP \cite{conf/fpga/DaiCWY16} finds the optimal number of PEs by modeling the runtime as the maximum of the time spent on interval loading, edge loading, and edge processing since those can be overlapped.
ForeGraph \cite{conf/fpga/DaiHCXWY17} extends this model by also including time to load intervals from other boards in their multi-FPGA setup and compares theoretical performance against other systems.
Zhang et al. specify a performance model that is a slight variation of the network model \cite{DBLP:conf/fpga/ZhangKL17}.
The memory system is represented by the packet size, packet overhead, bandwidths, and internal latency of memory.
An upper-bound performance model for vertex-centric graph processing on multiple FPGA systems is proposed by Engelhardt et al. \cite{DBLP:conf/fpl/EngelhardtHS18}.
The solution includes an architecture generator for multiple FPGAs with an application kernel and fitting dataset.



\subsubsection{Document}
Another early NRDS class is document (cf. \cref{fig:histogram}) with a focus on XML in the FPGA-accelerated literature. 
While JSON is the predominant document format in commercial document stores (cf. \cref{sec:document}), we did not find any solutions in the literature.
In the following, we will discuss solutions we found to the system aspects for the document NRDS class.

\paragraph{Operator}
In the literature, we found operators for XML parsing, document filtering, XPath evaluation, Twig, and XML projection which we subsequently discuss.

\labeltitle{XML parser}
Parsing of XML includes tasks like well-formed checking, schema validation, and tree construction.
In \cite{conf/fpga/DaiNZ10}, Dai et al. propose a solution that leverages recurring idioms in XML processing (one-to-one string match, one-to-many string membership test, one-to-many string search), and a speculative pipeline structure for tree construction skewed for common case high-throughput and edge case pipeline stalls. 
The FPGA is placed close to the network on a SmartNIC.
An alternative approach by Sidhu implements tree automata as a pair of a lexical and a tree automaton where the states of the lexical automaton form the transitions of the tree automaton \cite{conf/fpt/Sidhu13}.
%
Huang et al. propose a sliding window XML parsing accelerator \cite{HuangJWWP14}. 
They assume that the XML is valid and based on that can process multiple non-delimiter characters in one cycle.
Delimiter characters are still processed one by one.



\labeltitle{Filter}
Chalamalasetti et al. \cite{conf/ispass/ChalamalasettiMVWR12} provide an implementation of document filtering with scoring against topic profiles.
The work mainly improves the bloom filter design of \cite{conf/fpl/VanderbauwhedeAM09}.
The new design leverages multiple banks and reduces contention.



\labeltitle{XPath}
Mitra et al. propose a solution for XPath-based filtering of XML documents by mapping the XPath queries to regular expressions \cite{journals/corr/abs-0909-1781}.
These expressions are clustered by common profile prefixes and mapped to FPGA state machines (one per XPath).
A global stack is used for the inherent parent-child relationships.
%
In \cite{conf/reconfig/El-HassanI10}, publish-subscribe systems are extended to become an XML broker using an XPath processor with CAM.
The article also provides a hardware-based XML parser.
Moussalli et al. \cite{conf/hipeac/MoussalliSNT10} address the challenge of recursive XML filtering.
For that, each XPath is mapped into a stack whose width matches the XPath depths in bits and the height corresponds to the depth of the document.
Open tags are handled as push events and close tags as pop events.

\labeltitle{Twig}
The same authors propose an FPGA-based solution for twig matching on XML documents based on \cite{conf/hipeac/MoussalliSNT10} combined with dynamic programming \cite{conf/icde/MoussalliSNT11}.



\paragraph{Queries} 
For document stores, two different solutions for microsecond reprogrammable state machines as integral parts of string matching were proposed.
This does not allow query processing in itself but chaining of operators which can be used for query processing.

\labeltitle{Skeleton automata}
Teubner et al. propose the idea of skeleton automata as a fixed finite state automaton structure with parameterized transitions, allowing dynamic workload change in microseconds, instead of long (partial) reprogramming of the FPGA.

\labeltitle{Other}
The ZuXA system \cite{conf/www/LunterenEBCL04} implements a general programmable state machine with hash index on a rule table and a clustering scheme for very large automata that is applied as an XML acceleration engine.


\paragraph{CPU-FPGA collaboration}
We found three different collaboration schemes: socket, near-data, and PCIe.

\labeltitle{Socket}
Vanderbauwhede et al. address the challenge of power consumption of information filtering on streams of documents with a multi-FPGA setup \cite{conf/fpl/VanderbauwhedeAM09}.
A CPU places the document stream into main memory from where it is fetched by the FPGAs.
An on-chip BRAM Bloom filter is used to quickly discard irrelevant documents before the documents are matched against profiles stored in a hash table in on-board memory.


\labeltitle{Near-data}
The XLynx system \cite{journals/tods/TeubnerWN13} by Teubner et al. provides a solution for hybrid CPU-FPGA XQuery processing with dynamic XML projection.
The FPGA implements the same XML projector as in \cite{conf/sigmod/TeubnerWN12} placed into the data path between the document server and XQuery engine such that data is filtered before being queried, reducing load on the XQuery engine.


\labeltitle{PCIe}
In \cite{journals/sigarch/VanderbauwhedeFCM13}, previous work on document filtering \cite{conf/ispass/ChalamalasettiMVWR12, conf/fpl/VanderbauwhedeAM09} is advanced by embedding it into a hybrid CPU-FPGA system.
The CPU handles parsing the network document stream passing it to the FPGA as 64bit words using separator words between documents.
Additionally, the words are dictionary encoded. 


\subsubsection{Key-value} 
The most recent, NRDS-related research on FPGAs concerns key-value (cf. \cref{fig:histogram}), arguably the conceptually simplest class of NRDS.
We subsequently present the system aspect solutions for key-value stores.

\paragraph{Operator} 
Solutions are found for the externally visible insert and integral internal hash operator.

\labeltitle{Insert}
Liang et al. address the unpredictable insert performance of cuckoo hashing.
It avoids hash collisions by computing multiple hashes per key and reinserts one key-value pair when all hash positions are already occupied \cite{conf/fpl/LiangYKW16}.
The to-be-reinserted pair possibly triggers another pair to be reinserted into the hash table stalling naive pipelines.
The proposed solution splits up the pipeline into an insert and reinsert pipeline. 


\labeltitle{Hash}
Fast key-based value access requires reliable hash-functions like Murmurhash2\footnote{Murmurhash, visited 7/2020: \url{https://github.com/aappleby/smhasher}} missing on FPGAs.
Liu et al. contribute an implementation of Murmurhash2 on FPGA with different kernels for different key sizes that are applied through dynamic reconfiguration \cite{conf/hpcc/LiuZJJ19}.


\paragraph{Binary representation} The predominant binary representations are based on hash tables but there is also a CAM implementation.

\labeltitle{Hash table}
Istvan et al. leverage the FPGA's scalability and energy efficiency for a hash table implementation sustaining 10Gbps by implementing a sophisticated pipeline with concurrency control and separate key-hash and value store \cite{conf/hotcloud/BlottKLVBI13, conf/fpl/IstvanABV13}.
Collision handling is done with buckets by chaining fixed length pre-allocated memory regions for a tradeoff between probability of collisions and memory bandwidth.
Tong et al. present a hash table with one operation per clock cycle throughput \cite{conf/ipps/TongZP15}.
They propose two hash table access schemes.
The first one provides multiple slots for each hash value to reduce collisions where each operation scans all slots.
For bandwidth limited deployments, instead of scanning all slots, a second hash function decides the slot to work on similar to cuckoo hashing.
The tradeoff is again between probability of collisions and memory bandwidth.
FASTHash provides even higher throughput by processing $p$ queries in each cycle \cite{YangKSKP20}.
This is achieved by using $p$ parallel, data-independent PEs with eventual consistency of updates.
The hash table is split up into $p$ partitions each owned by one of the PEs.
Each partition is replicated to all PEs but only the PE that owns it inserts new values into it. 





\labeltitle{Content-addressable memory (CAM)}
Lockwood et al. propose a CAM-based, network-attached solution \cite{conf/hoti/LockwoodM15}.
They define an own message format for optimized hardware parsing of requests.
The key-value data is either stored in BRAM or DRAM while only BRAM guarantees low latency access and the value addresses are looked up with an emulated CAM.


\paragraph{Multi-tenancy}
Multi-tenancy is about performance and data isolation between multiple users working concurrently on a system.

\labeltitle{Token bucket}
Istvan et al. achieve this by defining traffic shapers with token buckets (from networking) and introducing tenant-specific registers for temporary query data \cite{conf/fpl/IstvanAS18}.


\paragraph{Scalability}
Data center level scalability is required by many applications using key-value stores.
There are solutions on data replication for increased read throughput and data partitioning to aid work distribution.

\labeltitle{Replication}
The Caribou system addresses scalability and fault-tolerance with data replication \cite{journals/pvldb/IstvanSA17}.
The FPGAs are deployed as a distributed storage layer in the storage nodes and allow for operator push-down (full scan and value predication) for near-data processing.
For scalability, key-value pairs are replicated between the nodes such that read requests can be served by any node in the storage layer.


\labeltitle{Partitioning}
FULL-KV \cite{QiuXLYLWYCGLS20} is a network-attached accelerator for CPU-FPGA hybrid systems, extending \cite{conf/fpt/QiuLXYW18} to two nodes.
The key-value store is partitioned to the nodes and requests are routed by a proxy.
BlueCache \cite{journals/pvldb/XuLJLHA16} acts as a caching layer of distributed network-attached FPGAs.
For operation with BlueCache, all application servers are equipped with a PCIe-attached FPGA that is connected to the other FPGAs via network and Flash for data storage.
Each CPU collects key-value store requests and passes them to its accelerator card in batches.
After parsing, requests are routed to the FPGAs containing the data and answered there.



\paragraph{Availability}
Availability is about fault tolerance meaning responsiveness even when single nodes fail.

\labeltitle{Replication}
Besides its contribution to scalability, the Caribou system \cite{journals/pvldb/IstvanSA17} provides availability through replication.
Writes are replicated from a master node to all other nodes in the storage layer.
Data is still available when one node fails.


\paragraph{Consistency}
For the consistency system aspect, there is one solution on isolation with MVCC, thus covering only one facet of consistency.

\labeltitle{Multi-version concurrency control (MVCC)}
Ren et al. define and implement MVCC support on top of \cite{conf/fpt/QiuLXYW18} by storing a version-value B-tree for every key \cite{conf/fpl/RenLSXQLYWYCH19}.
They also define atomic operations like compare and swap, compare and get, and predecessor version.


\paragraph{CPU-FPGA collaboration} 
We found near-data solutions with the FPGA between CPU and data and solutions based on direct memory access to expand value storage. 

\labeltitle{Near-data}
In \cite{journals/cal/LavasaniAC14}, the FPGA is used as an in-line accelerator for Memcached acceleration processing 96\% of the requests.
The CPU is only used as a fallback.
Based on \cite{conf/fpt/QiuLXYW18}, Xie et al. design an FPGA distributed memory proxy with four pipelined data paths and a pipelined consistent hashing processor \cite{conf/icfpt/XieQYW19}.
The orchestrator for the key-value store reaches up to 100Gbps.



\labeltitle{Direct memory access (DMA)}
Li et al. propose a network attached key-value store based on SmartNICs that accesses the main memory over PCIe DMA \cite{conf/sosp/LiRXLXPCZ17}.
%
The problems of the CPU cache hierarchy and FPGA low storage capacity are addressed by Qiu et al. \cite{conf/fpt/QiuLXYW18}.
Similar to \cite{conf/sosp/LiRXLXPCZ17}, the hash table is in on-board memory and the data in main memory (accessed with PCIe DMA).
The solution features novel memory allocation and fragmentation schemes.
%



\paragraph{Memory access}
For key-value stores, we found solutions for Bloom filters and Flash storage.

\labeltitle{Bloom filter}
Cho et al. propose a key-value store with cuckoo hashing and decoupled hash table and values \cite{conf/vlsi-dat/ChoC14}.
A Bloom filter is used to control hash table read access and thus reduce the amount of memory requests.


\labeltitle{Flash}
Flash storage was proposed as a viable storage medium for values by Blott et al. \cite{conf/hotstorage/BlottLKV15}.
Values are much larger in size than keys and are only very selectively accessed after their address has already been found over the hash table.
Thus, storing values in Flash storage allows much larger amounts of data to be stored.
Blott et al. scale out a Memcached server to 40 Terabytes of data this way \cite{conf/hotstorage/BlottLKV15}.
Similarly, BlueCache stores the hash table as an index cache in RAM and stores the corresponding values on the attached Flash storage \cite{journals/pvldb/XuLJLHA16}.


\paragraph{Performance model}
We found one solution modeling the performance bottleneck of the accelerator design.

\labeltitle{Bottleneck}
Qiu et al. examine the theoretical performance of their system based on the network performance as the bottleneck of the system \cite{QiuXLYLWYCGLS20}.
They split up their analysis by put and get operator.

%



\subsection{Synthesis and discussion of system aspects}
The system review in \cref{sec:background} resulted in a taxonomy of relevant FPGA and NRDS system aspects (cf. \cref{fig:combined}) that guided the literature analysis that addresses hypothesis (H2), \ie \emph{there are significant gaps in current research on non-relational FPGA acceleration} as set out in \cref{sub:methodology}.

To address (H2), we first answer question (a) What are the most relevant of the identified NRDS classes? by studying potential contributions chronologically.
We found that while document still plays a role in recent research efforts, current emphasis lies on accelerating graph and key-value.
This also supports the necessity of this work to consider FPGA-accelerated NRDS and their design decisions.

Secondly, we strive to answer questions (b) Are there any system aspects that are not yet covered by literature? and (c) Do existing approaches provide solutions to these topics? through the selection and summarizing of research literature.
\Cref{tab:implementations} sets the relevant FPGA and NRDS system aspects into context to the NRDS classes.
The focus on system aspects comes from our objective to guide practitioners and research towards FPGA-accelerated NRDS.

We found a large body of work on general graph operators (\eg BFS) and binary representation in the HPC domain.
While HPC does not focus on database processing, the work denotes a starting point for the graph database research.
Notably, when it comes to database-specific system aspects like queries (\ie GraphSoC \cite{conf/asap/Kapre15}), scalability (\ie most notably ForeGraph for multi-FPGA graph traversal \cite{conf/fpga/DaiHCXWY17}
), and consistency (\ie simple locking \cite{conf/fpga/MaZC17}) only few solutions are provided.
Due to the focus on FPGAs, there is an equally large body of work on design paradigms (\eg HitGraph \cite{conf/fccm/ZhouCP16, journals/tpds/ZhouKPSW19,conf/cf/ZhouKZP18}), CPU-FPGA collaboration (\eg most notably ExtraV \cite{DBLP:journals/pvldb/LeeKYCHNNJ17}), memory access, and performance models.
No solutions were found for availability, multi-tenancy, and security.

In the document class, there are only few solutions for functional system aspects.
Most notable is the XLynx system \cite{conf/sigmod/TeubnerWN12, journals/tods/TeubnerWN13}.
Non-functional system aspects are not covered.
FPGA system aspect solutions were only found for CPU-FPGA collaboration \cite{journals/tods/TeubnerWN13, conf/fpl/VanderbauwhedeAM09, journals/sigarch/VanderbauwhedeFCM13}.
All other categories remain without a solution.

For key-value stores, important functional topics like operators and binary representation are covered (\eg hash tables \cite{conf/fpl/IstvanABV13,conf/ipps/TongZP15, YangKSKP20} and CAM \cite{conf/hoti/LockwoodM15}).
Queries and sophisticated design paradigms are not applicable to key-value since there are only CRUD operators.
The non-functional system aspects scalability, availability, and multi-tenancy are mainly studied in the Caribou system \cite{conf/fpl/IstvanAS18,journals/pvldb/IstvanSA17}.
Further solutions are provided for consistency (\ie isolation with MVCC \cite{conf/fpl/RenLSXQLYWYCH19}) as well as CPU-FPGA collaboration, memory access, and one performance model.
We found no solutions for security.

In summary, we only found FPGA-accelerated NRDS for key-value databases, while in the other NRDS classes, solutions for functional system aspects are available.
For NRDS, these solutions would have to be adapted to databases to be usable.
There were only few non-functional solutions provided.

\section{FPGA-accelerated NRDS}
\label{sec:practitioners_guide}

This section gives several perspectives on NRDS class commonalities, addressing hypothesis (H3), \ie \emph{there are patterns guiding the design of an FPGA-accelerated non-relational database system}.
We revisit existing practical FPGA-accelerated NRDS solutions \cite{capiSnap, redisFpga, cassandraFpga} (cf. \cref{sec:background}) and the current state in research (cf. \cref{sec:literature}) and discuss overarching patterns that we found 
along four questions on building an FPGA-accelerated NRDS starting from the idea or need of accelerating a system, over its design and implementation, and finally the evaluation of the accelerator's impact:
\begin{enumerate}[label=Q\arabic*]
    \item Which problems are FPGAs able to solve? (accelerator task: \cref{sec:task})
    \item Where to put the FPGA(s)? (accelerator placement: \cref{sec:placement})
    \item How to implement NRDS operators? (accelerator design: \cref{sec:design})
	\item How to measure improvements of the NRDS? (accelerator justification: \cref{sec:justification})
\end{enumerate}

Subsequently, we give answers to questions Q1--4 in the form of a practitioners guide.


\subsection{Accelerator task}
\label{sec:task}


When dealing with a concrete system implementation, either an accelerator is added to an existing system or becomes relevant to the design of a new one.
However, in both cases one has to decide whether an FPGA is suitable.
As a first step, we answer question Q1 (Which problems are FPGAs able to solve?) by summarizing the problems typically solved by FPGAs in the different NRDS classes and differentiating task categories that are well suited to FPGAs.
Notice, however, that in this survey we cannot represent all possible motivations and tasks that are met by practitioners in practice.

\paragraph{Graph}
For graph processing, FPGAs are mainly used to offload operators because of inefficiencies in the cache hierarchy and coarse-grained memory access of CPUs resulting from the irregular memory access patterns of graph workloads (cf. \cite{conf/ipps/AttiaJTJZ14, conf/fpga/DaiHCXWY17, journals/tpds/ZhouKPSW19}).
An FPGA has more control over its memory access and thus helps to alleviate the problem of irregular memory accesses through custom memory controller design and full control of data placement in on-chip memory (cf. \cref{sec:memory_access}).

\paragraph{Document}
FPGAs for document processing are mainly used as bandwidth amplifiers for the CPU. 
Much of document processing is parsing and filtering with large data movement costs putting heavy load on the CPU even tough a lot of the data is discarded and pollutes the cache hierarchy.
Thus, FPGAs can be used as a flexible stream processing accelerator in the data path to the memory (\eg \cite{conf/ispass/ChalamalasettiMVWR12}), disk, or network (\eg \cite{conf/fpga/DaiNZ10}).
Sometimes, however, the CPU is completely bypassed and the FPGA is used as a standalone accelerator in the network (\eg \cite{conf/sigmod/TeubnerWN12}).

\paragraph{Key-value}
For key-value stores, the literature mainly shows two schemes for motivation of FPGA usage. 
Either a single function instrumental to key-value stores is accelerated (\eg insertion \cite{conf/fpl/LiangYKW16} or hashing \cite{conf/hpcc/LiuZJJ19}) because the CPU does not meet the latency requirements, or building a full system is motivated by large roundtrip latencies of CPUs from the network through the operating system network stack and back (\eg \cite{journals/pvldb/IstvanSA17, conf/sosp/LiRXLXPCZ17, QiuXLYLWYCGLS20}).

\paragraph{Summary -- task categories}
\begin{figure}[bt]
    \centering
    \includegraphics[width=\linewidth]{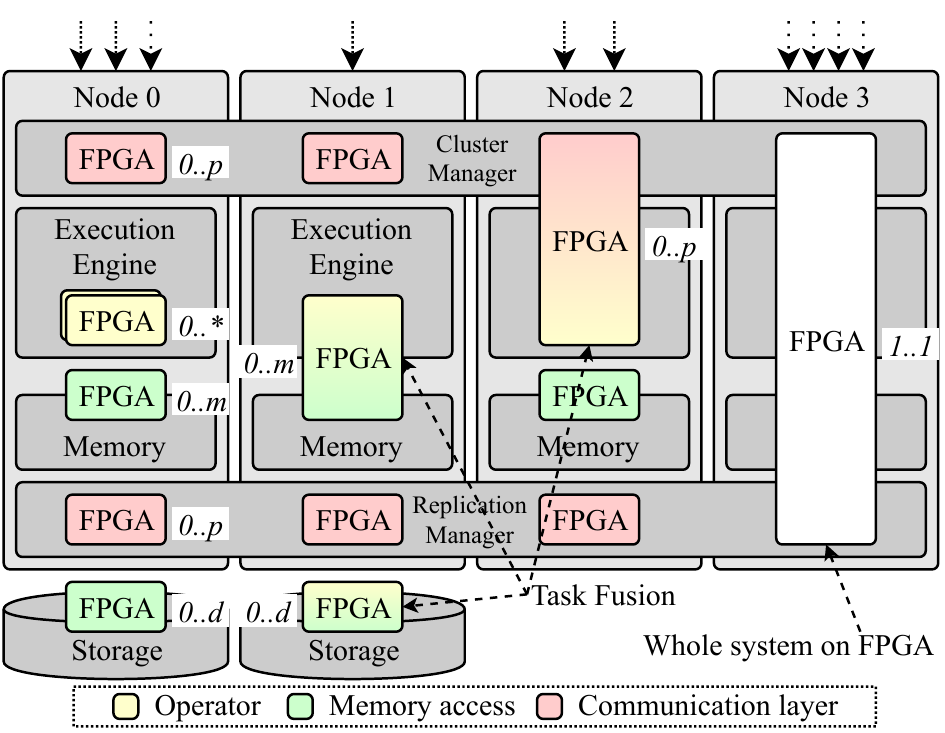}
    \caption{Potential FPGA tasks in NRDS}
    \label{fig:tasks}
\end{figure}
The two biggest problems that FPGAs solve for the different NRDS classes are:
%
\begin{inparaenum}[\it (1)]
    \item data movement from peripherals (\eg network or disk) to the CPU and
    \item memory access inefficiencies by the fixed CPU architecture.
\end{inparaenum}
\Cref{fig:tasks} shows a simplified version of the shared system architecture from \cref{fig:combined}.
The tasks that FPGAs might take over in an NRDS to address these problems fall into three categories: operator acceleration, data access acceleration, and communication layer.

Operator acceleration focuses on improving performance for one or multiple operators of the NRDS class.
A lot of the literature focuses on this task category with implementations of specific operators.

Particularly interesting for NRDS are accelerator implementations in the context of hybrid CPU-FPGA systems (\eg \cite{conf/fccm/BondhugulaDDFWSS06, journals/iet-cdt/JagadeeshSL11, conf/fpl/VanderbauwhedeAM09, journals/sigarch/VanderbauwhedeFCM13, DBLP:conf/fccm/WangHN19}).
In data-centric applications, the FPGA can take pressure off the CPU with data access acceleration.
One already existing example is graph virtualization, where non application-specific access patterns are accelerated on an FPGA near storage \cite{DBLP:journals/pvldb/LeeKYCHNNJ17}.

Placing FPGAs in the communication layer is another promising option.
One example is a proxy layer for key-value stores where the request router (cf. \cref{fig:combined}) is placed in an FPGA outside the other nodes which routes traffic to the correct CPU nodes \cite{conf/icfpt/XieQYW19}.

On node 1 and 2 in \cref{fig:tasks} we show possible combined acceleration of tasks that we call \emph{task fusion}.
Task fusion is possible when the resources on the FPGA fit both tasks and should be done if more than one task benefits from FPGA acceleration.
The tasks combined on one FPGA may even accelerate the overall system more than if they would be accelerated separately because data movement costs are reduced.

In the relational database literature we found examples of operator push down where operator acceleration and data access are fused into one (\eg \cite{Francisco11, journals/pvldb/WoodsIA14}).
This worked especially well for filter operators pushed to the FPGA that reduced the amount of data communicated to the CPU.
This kind of accelerators use their close proximity to memory to reduce the data load on the CPU.
The newly presented Enzian system \cite{conf/cidr/AlonsoRCEKKSW20} also enables this with a cache-coherent attachment of the FPGA and opens up questions of near-data processing where the FPGA is used for on-the-fly conversion of binary representation.
One prominent example of fusing operators with the communication layer at data center scale is the Microsoft Catapult project \cite{conf/isca/PutnamCCCCDEFGGHHHHKLLPPSTXB14}.
There, servers push document classification workloads to a communication layer of multiple FPGAs connected to each other.
Another example from the key-value literature is \cite{conf/fpt/QiuLXYW18}.
They use the FPGA as an entry point from the network and do pre-processing of queries in the FPGA while -- due to size -- storing the actual data values in the memory of the CPU.

In extreme cases, one to all nodes in the NRDS cluster can be mapped to FPGAs (node 3) (\eg \cite{journals/pvldb/IstvanSA17}).
This works well for key-value systems (and might work for wide-column systems) since the overall system is simple enough, and also works for providing standalone services on FPGAs (\eg \cite{journals/tods/TeubnerWN13}).
In this case the FPGA has to be network-attached which saves a lot of overhead by not going through the CPU cache hierarchy and operating system network stack.

To aid the decision of how many FPGAs to put into a system we added cardinalities to each possible FPGA task (\cref{fig:tasks}).
In the confines that are set by the system hardware, independent FPGAs could be added for different tasks.
One FPGA can be added for each network port ($p$ is number of network ports), memory subsystem, and disk ($m$ is number of memory subsystems; $d$ is number of disks) in the system.
For operators, there is no such restriction. 
There can be as many FPGAs as fit into the hardware system.
Insight 1 answers question Q1 (Which problems are FPGAs able to solve?):
\begin{insight}
    There are three accelerator task categories (operator, data access, and communication layer  acceleration) that FPGAs are currently well suited for.
\end{insight}

\subsection{Accelerator placement}
\label{sec:placement}


After deciding for which task the FPGA accelerator is used in the system, in this section, we discuss FPGA placement patterns in the context of one cluster node. 
In the literature, we discovered four FPGA placement patterns which we discuss before we show how to chose a placement based on the task (cf. \cref{sec:task}) and properties of the workload. This answers question Q2 (Where to put the FPGA(s)?).

\paragraph{Placement patterns}
\label{sec:placement_patterns}

We differentiate between the main memory of the overall system (SysRAM), attached to the CPU, and RAM directly attached to the FPGA on the board (FRAM).
\begin{figure}[bt]
    \centering
    \includegraphics[width=.6\linewidth]{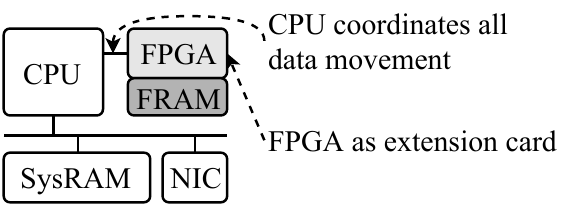}
    \caption{Placement pattern: \emph{Offload}}
    \label{fig:placement_offload}
\end{figure}

The \emph{offload} accelerator placement (\cref{fig:placement_offload}) is defined in being attached only to the CPU (\eg over PCIe) and an on-board FRAM much smaller than SysRAM. 
Input data is directly written to FRAM by the CPU and execution is triggered by the CPU. 
The FPGA works on the input data in FRAM, and the results are transferred by the CPU to the SysRAM on notification of the CPU by the FPGA. 
This placement only allows master-slave setups, and thus can introduce a lot of overhead for acceleration because the FPGA cannot move data in the system on its own, and CPU cycles are wasted on data movement.
The FPGA-accelerated in-memory database survey \cite{journals/vldb/FangMHLH20} shows this placement as \emph{IO-attached accelerator}.
\begin{figure}[bt]
    \centering
    \includegraphics[width=.6\linewidth]{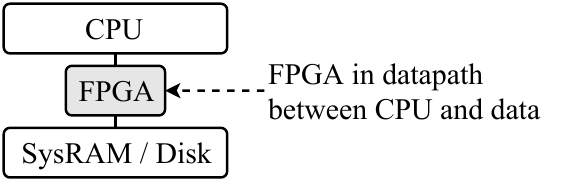}
    \caption{Placement pattern: \emph{Near-data}}
    \label{fig:placement_neardata}
\end{figure}

The second placement we found in the literature is the \emph{near-data} placement (\cref{fig:placement_neardata}). 
It is defined by the way the FPGA is inserted into the data path between the CPU and the SysRAM or disk.
In this way, the FPGA provides an interface to the CPU to interact with the underlying resource.
In a more restricted way, \cite{journals/vldb/FangMHLH20} define this as a \emph{bandwidth amplifier} which decompresses data, however this placement can accelerate more workloads than just decompression, \eg filtering and binary representation conversion.
In \cite{conf/sigmod/MuellerT09} there is a similar placement where the FPGA is placed in the data path between the disk and the CPU.
\begin{figure}[bt]
    \centering
    \includegraphics[width=.6\linewidth]{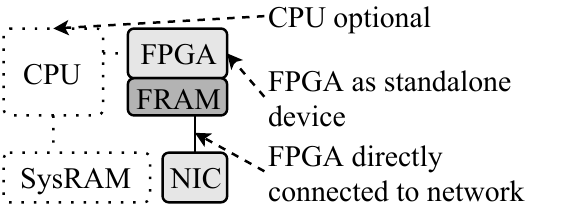}
    \caption{Placement pattern: \emph{SmartNIC}}
    \label{fig:placement_smartnic}
\end{figure}

\Cref{fig:placement_smartnic} shows the \emph{SmartNIC} placement where the FPGA is directly attached to the network interface controller (NIC). 
This placement option may even completely eliminate the CPU in the system if there are no tasks besides what is implemented on the FPGA. 
The SmartNIC placement optimizes for low latency of the overall system by saving multiple round trips through the operating system kernel on the CPU. 
\begin{figure}[bt]
    \centering
    \includegraphics[width=.6\linewidth]{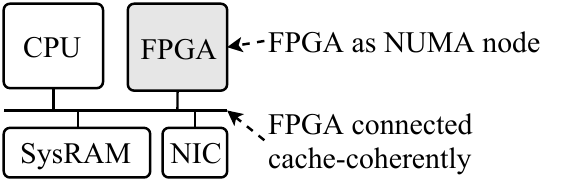}
    \caption{Placement pattern: \emph{Socket}}
    \label{fig:placement_socket}
\end{figure}
\begin{figure*}[t]
    \centering
    \includegraphics[width=.9\linewidth]{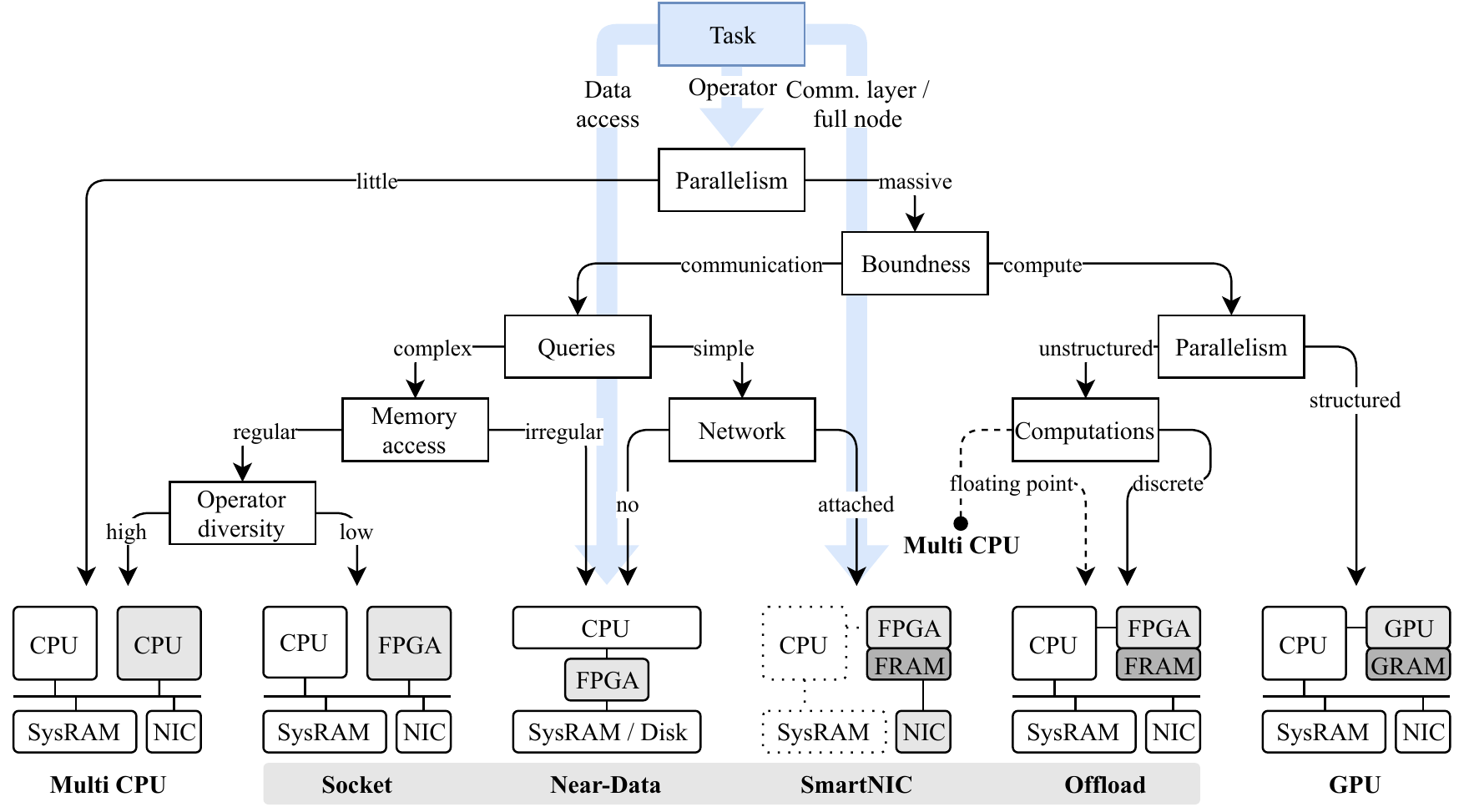}
    \caption{Accelerator placement decision tree}
    \label{fig:decision_placement}
\end{figure*}

In emerging systems (\eg \cite{conf/cidr/AlonsoRCEKKSW20}), the FPGA may be placed as a \emph{socket} (\cref{fig:placement_socket}) at least conceptually with a cache coherent access to SysRAM. 
The three previously discussed placements can be represented with the FPGA being a socket with little overhead. 
However, the socket placement also enables new work distribution strategies where the CPU does not have to coordinate execution of the FPGA and data movement.
In \cite{journals/vldb/FangMHLH20} this placement option is called \emph{co-processor}.

\paragraph{Placement decision}

FPGA placements discussed in the literature can be reduced to the four fundamental patterns from \cref{sec:placement_patterns}.
Based on these, \cref{fig:decision_placement} shows a decision tree guiding the practitioner towards choosing an accelerator placement pattern depending on task and workload properties.
Additionally, we added CPUs and GPUs as alternative accelerator options. 

For tasks that incur large data movement overheads from either the memory, disk, or network we have introduced shortcuts (shown as blue arrows) to the near-data and SmartNIC placements respectively.

For workloads that do not exhibit massive parallelization opportunities we do not see much potential in applying an accelerator.
Thus, this leads to adding more CPUs to the system or alternatively adding more nodes to the cluster.
For compute-bound problems with structured parallelism, meaning large numbers of homogeneous threads running in parallel, we would chose GPUs over FPGAs because they are specifically made to handle these workloads \cite{journals/micro/LindholmNOM08}. 
Similarly, for tasks with heavy reliance on unstructured floating-point operations, we would most of the time advise against using an FPGA as an accelerator because the DSP floating point units on the FPGA will quickly become the bottleneck.

For compute-bound problem instances that are not better suited to GPUs or multi-CPU, an \emph{offload} accelerator approach is chosen.
The data movement is a big source of overhead in this placement model such that it only works for compute-bound problems where data movement costs, on the slow link between CPU and FPGA, are negligible compared to the duration of the computation. 
The offload accelerator approach is implemented by most of the graph literature (\eg \cite{conf/fpga/DaiHCXWY17, DBLP:conf/dac/FinnertySLL19, journals/tpds/ZhouKPSW19}). 
While we think that the offload pattern could be a viable option for database systems, we focus on more significant improvements through acceleration. 

In the category of memory-bound problems, we differentiate first between workloads with simple operators (\eg lookup) and second workloads with complex operators (\eg graph traversal).
Simple queries are defined as a combination of few simple operators with few predicate expressions. 
This category includes key-value and wide-column stores and can include document and graph stores in certain scenarios (\eg data provider for graph neural networks). 
If the database system is network-attached, we choose a \emph{SmartNIC}.
This placement was also found to be efficient in related data processing domains like data-intensive messaging \cite{DBLP:conf/middleware/Ritter17, DBLP:conf/debs/RitterDMR17}.
If the database system is only part of a larger architecture and not network-attached, we chose the \emph{near-data} approach.
This placement option is also chosen when there are complex queries with irregular memory accesses (\eg in graph traversal).
We think that the \emph{socket} placement will benefit \gls{fpga} accelerated systems especially in the database context.

As shown in \cref{fig:decision_placement}, adding an FPGA to the system is not always the best strategy to improving performance of a system. 
Moreover, the traditional approach of placing an accelerator in a system as an offload accelerator is not the best option for database systems.

\begin{table*}[t]
\scriptsize
\centering
\begin{tabular}{l l |c|l l l l|c l c}
    NRDS class & System identifier & Task & Queries & Mem. access & Op. div. & NRDS & Decision & Correct \\
    \hline
    \hline
    Graph & CAPI SNAP \cite{capiSnap} \faWrench & D & complex & irregular & n/a & \faThumbsOUp & Near-data & \faThumbsOUp \\
    & HAGAR \cite{conf/fpl/MencerHH02} & O & complex & irregular & n/a & \faThumbsDown & Near-data & \faThumbsDown (Offload) \\
    & GraphStep \cite{conf/fccm/DeLorimierKMRERUKD06} & CO & complex & irregular & n/a & \faThumbsDown & SmartNIC & \faThumbsOUp \\
    & CyGraph \cite{conf/ipps/AttiaJTJZ14} & O & complex & irregular & n/a & \faThumbsDown & Near-data & \faThumbsOUp \\ 
    & TorusBFS \cite{journals/iracst/LeiRG15} & CO & complex & irregular & n/a & \faThumbsDown & SmartNIC & \faThumbsOUp \\
    & FPGP \cite{conf/fpga/DaiCWY16} & O & complex & irregular & n/a & \faThumbsDown & Near-data & \faThumbsDown (Offload) \\
    & ForeGraph \cite{conf/fpga/DaiHCXWY17} & CO & complex & irregular & n/a & \faThumbsDown & SmartNIC & \faThumbsOUp \\
    & ExtraV \cite{DBLP:journals/pvldb/LeeKYCHNNJ17} & D & complex & irregular & n/a & \faThumbsDown & Near-data & \faThumbsOUp \\
    & Dr. BFS \cite{DBLP:conf/dac/FinnertySLL19} & O & complex & irregular & n/a & \faThumbsDown & Near-data & \faThumbsDown (Offload) \\
    & FabGraph \cite{DBLP:conf/fpga/ShaoLHL019} & O & complex & irregular & n/a & \faThumbsDown & Near-data & \faThumbsDown (Offload) \\
    & HitGraph \cite{journals/tpds/ZhouKPSW19} & O & complex & irregular & n/a & \faThumbsDown & Near-data & \faThumbsDown (Offload) \\
    \hline
    Document & ZuXA \cite{conf/www/LunterenEBCL04} & O & complex & regular & low & \faThumbsDown & Socket & \faThumbsDown (Offload) \\
    & XLynx \cite{journals/tods/TeubnerWN13} & CO & complex & regular & low & \faThumbsDown & SmartNIC & \faThumbsOUp \\
    \hline
    Key-value & Algo-Logic \cite{redisFpga} \faWrench & C & simple & n/a & n/a & \faThumbsOUp & SmartNIC & \faThumbsOUp \\ 
    & BlueCache \cite{journals/pvldb/XuLJLHA16} & D & simple & n/a & n/a & \faThumbsOUp & Near-data & \faThumbsOUp \\
    & Caribou \cite{journals/pvldb/IstvanSA17} & F & simple & n/a & n/a & \faThumbsOUp & SmartNIC & \faThumbsOUp \\
    & KV-Direct \cite{conf/sosp/LiRXLXPCZ17} & F & simple & n/a & n/a & \faThumbsOUp & SmartNIC & \faThumbsOUp \\
    & FULL-KV \cite{QiuXLYLWYCGLS20} & F & simple & n/a & n/a & \faThumbsOUp & SmartNIC & \faThumbsOUp \\
    \hline
    Wide-column & rENIAC \cite{cassandraFpga} \faWrench & C & simple & n/a & n/a & \faThumbsOUp & SmartNIC & \faThumbsOUp \\
\end{tabular}
\medskip
\begin{tablenotes}
    \centering
    \item[*] \faWrench: commercially available; task options (D: data access, O: operator, C: communication layer, F: full system), combinations are permitted; n/a: ``not applicable''; Mem. access: Memory access, Op. div.: Operator diversity; \faThumbsOUp: yes, \faThumbsDown: no
\end{tablenotes}
\caption{Accelerator decision validation for systems in the literature (all systems are communication bound)}
\label{tab:decision}
\end{table*}
\Cref{tab:decision} shows all systems found in the system review and literature analysis and how the placement decisions look like.
We exclude the systems describing frameworks without a description of how they are deployed (\ie GraphGen \cite{DBLP:conf/fccm/NurvitadhiWWHNH14}, GraVF \cite{DBLP:conf/fpl/EngelhardtS16}, and GraphOps \cite{DBLP:conf/fpga/OguntebiO16}).
For tasks that have a quick path to the placement patterns, the decision from the decision tree is always correct.
This includes the commercial solutions from the system review, \eg CAPI SNAP \cite{capiSnap} which uses near-data deployment of FPGAs for graph workloads.
For pure operator acceleration we propose near-data placement for graph and socket placement for document workloads.
This is only implemented by CyGraph indicating a lack of consideration for the whole system and therefore data movement in the offload placement pattern.
The offload placement pattern can be seen as the initial step an inexperienced practitioner might take, since it trades off performance against easy system integration.


\paragraph{Summary -- accelerator placement}
In this section we first showed how FPGAs can be attached to the other hardware components in a system.
Especially compared to a CPU, FPGAs can be placed close to the data, whether in memory, disk, or network.
%
\begin{insight}
    There are four fundamental patterns of FPGA placement (offload, SmartNIC, near-data, and socket).
\end{insight}

Thereafter, we established a decision tree guiding the practitioner from the tasks (cf. \cref{sec:task}) and the characteristics of the operators towards choosing a placement pattern.
We validated this decision tree with the systems found in the system review and literature analysis by comparing our and their placement decision.
\begin{insight}
    The accelerator task in combination with the characteristics of operators of the NRDS class are sufficient to decide the FPGA placement.
\end{insight}

\subsection{Accelerator design}
\label{sec:design}
Now that we know what the FPGA should do and where it is placed in the system we need to answer question Q3 (How to implement NRDS operators?).
Although implementation details largely depend on specific algorithms and data structures of the NRDS classes, we found patterns for two critical accelerator design considerations.
In the following we introduce operator switching strategies and memory access optimization patterns common for all NRDS classes on FPGAs.

\paragraph{Operator switching strategies}
\label{sec:dynamic_queries}
\begin{figure}[bt]
    \centering
    \includegraphics[width=\linewidth]{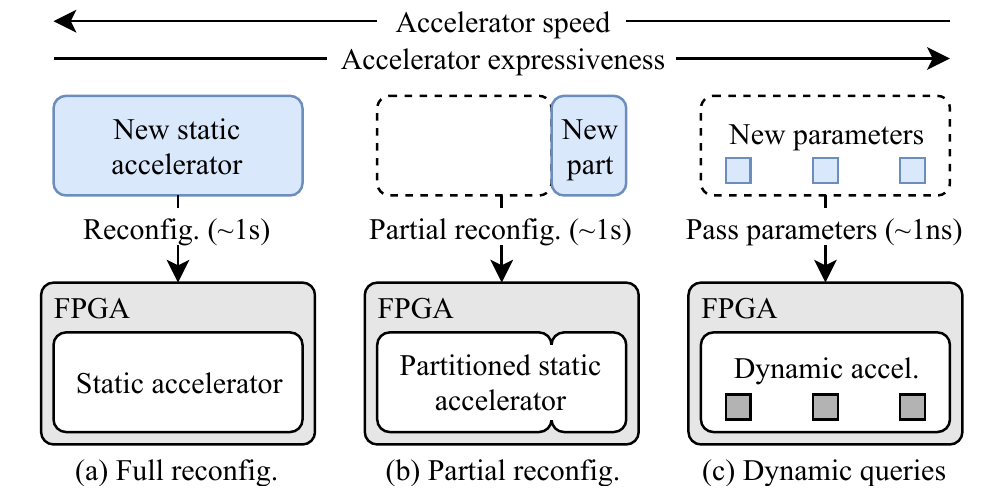}
    \caption{Strategies for operator switching}
    \label{fig:dynamic_queries}
\end{figure}
In the literature analysis we saw little about accelerators that are able to switch to different operators without compiling a new accelerator each time.
However, it is required of a NRDS accelerator to process multiple different operators in parallel and in very quick succession on multiple different datasets in memory, since it is not sufficient for an accelerator to improve the performance of one infrequently used operator to offset added cost and complexity.
This is easy to achieve on instruction-based architectures, like CPUs, since their programs are easily switched out but difficult to achieve on FPGAs since they cannot switch their architecture without significant overhead.
This is shown in \cref{fig:dynamic_queries}(a) as \emph{full reconfiguration} where an operator switch takes seconds \cite{journals/trets/PapadimitriouDH11}.

To alleviate the overhead of full reconfiguration, the relational database community pursued \emph{partial reconfiguration} (\cref{fig:dynamic_queries}(b)) where only parts of the accelerator architecture are switched (\eg \cite{conf/fccm/DennlZT13, conf/heart/ManevVKK19, conf/fccm/OwaidaSKA17, DBLP:journals/mam/WernerHPG17, journals/trets/ZienerBBDMSTVW16}).
While this works for coarse-grained functionality switching during runtime, the part that is reconfigured still is unavailable for seconds. 

Thus, we advocate for a more elegant and expressive solution in what we call \emph{dynamic queries} (\cref{fig:dynamic_queries}(c)).
The dynamic queries switching strategy is based on a dynamic, parameterized or instructable, domain-specific accelerator (\eg \cite{conf/fccm/IstvanSA16, conf/asap/Kapre15, conf/www/LunterenEBCL04, conf/IEEEpact/SukhwaniMTDIBDA12, conf/sigmod/TeubnerWN12, journals/tods/TeubnerWN13}) that allows to process multiple different operators in parallel and with only nanosecond switching delay on multiple datasets by just passing new parameters or instructions.
The added accelerator expressiveness might come at the loss of some accelerator speed \cite{journals/it/Teubner17} but saves magnitudes in operator switching delay.
The difficulty of designing such an accelerator lies in finding abstractions of high generality without introducing much overhead that slows down performance.
The examples we found above focus on a small set of domain-specific primitives that are combined into a dynamic accelerator.

Although we did not find references for it in the literature, the operator switching strategies can be applied in combination.
Depending on the task it might be beneficial to partially or even fully reconfigure the FPGA as long as it is done at a frequency low enough (\eg if lasting workload changes are detected).

\paragraph{Memory access optimization patterns}
\label{sec:memory_access}
\begin{figure}[t]
    \centering
    \includegraphics[width=\linewidth]{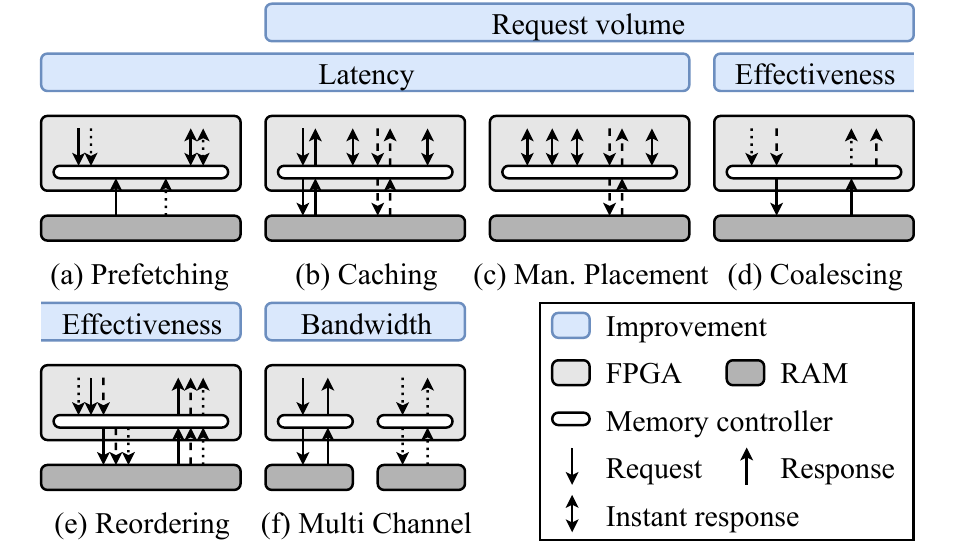}
    \caption{Memory access optimization patterns}
    \label{fig:memory_optimizations}
\end{figure}
For NRDS, memory accesses are one of the most instrumental challenges to good performance. 
We found six memory access optimization patterns (\cref{fig:memory_optimizations}) in the literature that are applicable to all NRDS classes.
Each pattern is implemented in the memory controller (endpoint to memory on the FPGA) of the accelerator design and is independent of the accelerators algorithms and data structures.
The memory controllers performance can be improved along four axes: by reducing \emph{latency} of accesses, reducing the number of accesses (\emph{request volume}), enhancing the amount of effective data per access (\emph{effectiveness}), and increasing raw memory \emph{bandwidth}.
In the following we introduce these six combinable patterns.


\labeltitle{Prefetching (cf. \cref{fig:memory_optimizations}(a))}
Prefetching is a technique to hide memory access latency that starves processing of input data.
Therefore, memory requests are issued before the data is processed to overlap computation and loading of data if the access locations are known beforehand. 
By the time the data is processed, it already resides on chip to be consumed.
One example is partitioning the data and overlapping partition loading with partition processing \cite{DBLP:conf/fpga/ShaoLHL019, DBLP:conf/reconfig/ZhouCP15}.
Another is issuing large amounts of non-blocking memory requests such that there is always data to process \cite{DBLP:conf/fpl/BetkaouiWTL12}.

\labeltitle{Caching (cf. \cref{fig:memory_optimizations}(b))}
Caching  (or automatic data placement) reduces high-bandwidth utilization by storing accessed values in on-chip memory (cache).
If the cache is full, there is an automated policy (\eg least recently used) replacing values in the cache with new ones \cite{JainL19}.
If a value in the cache is requested, it is instantly served from on-chip memory, but this only works well for workloads with strong temporal locality.
One example is multi-level caching in \cite{DBLP:conf/fpga/ShaoLHL019}.
In \cite{DBLP:conf/fccm/WangHN19}, Wang et al. combine caching with the before mentioned reordering to increase the spatial locality of memory accesses.

\labeltitle{Manual data placement (cf. \cref{fig:memory_optimizations}(c))}
FPGAs provide the practitioner with full control over what data resides in quickly accessible on-chip memory not only custom caching techniques can be employed but critical data can be permanently placed on the FPGA.
This may be done for frequently accessed data structures critical to the performance of the accelerator (\eg \cite{DBLP:conf/fpl/BetkaouiWTL12, conf/fpl/LiangYKW16, DBLP:conf/fpga/ZhangKL17, DBLP:conf/fpga/ZhangL18}).
Another example is storing a highly efficient data structure like a bloom filter that allows filtering of memory accesses for presence of values in a dataset (\eg \cite{conf/fpt/BecherZMT15, conf/vlsi-dat/ChoC14, conf/fpl/VanderbauwhedeAM09}).

\labeltitle{Coalescing (cf. \cref{fig:memory_optimizations}(d))}
Coalescing means merging multiple memory requests of single data items into one memory access (\eg \cite{conf/fpga/KhoramZSL18}).
Since modern DRAM operates on rows of memory that are in the kilobyte range, accessing single data items is wasteful.
If the workload exhibits strong spatial and temporal locality, coalescing can reduce the number of memory accesses per single data item and thus decrease the request volume by simultaneously increasing the effectiveness of each memory access.
In \cite{DBLP:conf/dac/FinnertySLL19} this is done by having many more compute units than data access units that issue many accesses enabling data access units to coalesce some of the accesses.
Another example is combining write requests before they are written to memory and thus reducing the overall number of writes \cite{conf/ipps/ZhouCP15, conf/fccm/ZhouCP16}.

\labeltitle{Reordering (cf. \cref{fig:memory_optimizations}(e))} 
Usable memory bandwidth suffers from irregular accesses.
Reordering of memory requests can improve upon this if the workload exhibits spatial and temporal access locality.
This can be done online (\eg \cite{DBLP:conf/islped/YanHLALMDYZF019}), at the cost of increased latency, or offline (\eg \cite{DBLP:conf/reconfig/ZhouCP15, conf/fccm/ZhouCP16}), if there is a correlation between data and memory access order.

\labeltitle{Multi Channel (cf. \cref{fig:memory_optimizations}(f))}
If multiple memory channels are available, the memory bandwidth can be increased by distributing memory accesses over those channels.
This is a meta-pattern that can be combined with any of the aforementioned patterns.
One example of using multiple channels is placing different data structures on different channels \cite{journals/tcas/LeiDLX16, DBLP:journals/ieiceee/NiDZLW14}.

\paragraph{Summary -- accelerator design}
While there was no focus on operator switching in the HPC-motivated literature, there has been some work on the topic especially in the document class.
We see it as a crucial consideration in FPGA-accelerated NRDS.
Furthermore, the memory access optimization patterns are especially important to NRDS since memory access acceleration is one of the big motivations to use FPGAs. 
Thus, we conclude with the following insight:
\begin{insight}
    There are three operator switching strategies and six memory access optimization patterns guiding the development of accelerators for NRDS.
\end{insight}

\subsection{Justification}
\label{sec:justification}
%


In an FPGA-accelerated system, the FPGA's improvement on performance must not be evaluated on the accelerated part of the workload in isolation but the whole system.
An FPGA introduces a new component into the hardware system that entails costs having to be justified by the performance improvement. 
Costs occur in the form of cost of ownership (which might benefit from FPGA energy efficiency) but also cost of programming and operating a whole new hardware architecture. 
While the decision if the performance improvement outweighs the cost lies in the judgement of the practitioner, in this section we will introduce measuring performance improvement with benchmarking and performance models for FPGA-accelerated NRDS.
Therewith we answer question Q4 (How to measure improvements of the NRDS system?). 

\paragraph{Benchmarks}
As a guideline to good benchmarking we follow the four key criteria for domain-specific benchmarks from \cite{gray1993benchmark}. 
Benchmarks should be easy to understand (\emph{simple}) and scale from small to powerful systems in the present and towards the future (\emph{scalable}). 
However, the two criteria most critical to FPGA-accelerated NRDS are that benchmarks are \emph{portable} and \emph{relevant} which we discuss in the following.

Benchmarking only works well when either comparing different systems with the same program or different programs with the same underlying system. 
Either the program or the system as variables have to be fixed for comparability. 
This especially poses a big problem for benchmarking on FPGAs since different FPGAs have very different specifications, and implementations are often tuned to one specific board. 
We did not find any solutions for this problem in literature.

\begin{table}[bt]
\scriptsize
\centering
\begin{tabular}{l l c c r r r}
 Domain & Name & Dataset & Workload & \#Refs \\
 \hline
 \hline
 Graph & graph500 & \faThumbsOUp & \faThumbsOUp & 1 \\
  & live-journal & \faThumbsOUp & \faThumbsDown & 7 \\
  & rmat & \faThumbsOUp & \faThumbsDown & 4 \\
  & pokec & \faThumbsOUp & \faThumbsDown & 4 \\
  & orkut & \faThumbsOUp & \faThumbsDown & 3 \\
  & twitter & \faThumbsOUp & \faThumbsDown & 3 \\
  & wiki-talk & \faThumbsOUp & \faThumbsDown & 2 \\
  & indochina & \faThumbsOUp & \faThumbsDown & 2 \\
  & flickr & \faThumbsOUp & \faThumbsDown & 2 \\
  & roadnet-ca & \faThumbsOUp & \faThumbsDown & 2 \\
 \hline
 Doc. & xmark & \faThumbsOUp & \faThumbsOUp & 4 \\
  & trec aquaint & \faThumbsOUp & \faThumbsDown & 3 \\
  & toxgene & \faThumbsOUp & \faThumbsDown & 3 \\
  & yfilter & \faThumbsDown & \faThumbsOUp & 3 \\
\end{tabular}
\medskip
\begin{tablenotes}
    \centering
    \item[*] \faThumbsOUp: yes, \faThumbsDown: no
\end{tablenotes}
\caption{Benchmark statistics}
\label{tab:benchmark}
\end{table}

Regarding relevance, we found different datasets and workloads listed in \cref{tab:benchmark}.
Excluding ``indochina'', all non-synthetic graph datasets can also be found in the Stanford Network Analysis Project (SNAP) graph collection\footnote{SNAP dataset collection, visited 7/2020: \url{https://snap.stanford.edu/data/index.html}}.
Considering the overall number of articles we found by NRDS class, it does not seem as if any workload or dataset is established especially when compared to \eg TPC in the database literature.
For key-value systems, benchmarking is currently done in literature as a sequence of a subset of the three basic API functions (\ie get, put, delete) without a standardized dataset.
However, a well-formed benchmark establishes not only datasets and workloads but all of the following artifacts:
\begin{inparaenum}[\it (1)]
    \item Workloads (\eg for graph: BFS, shortest path, weakly connected components)
    \item Datasets (\eg for graph: twitter, rmat, live-journal)
    \item Domain-specific performance measures (\eg for graph: traversed edges per second (TEPS))
    \item Benchmark (reference) implementation details.
\end{inparaenum}
Thus, current performance measurements not only lack relevance and artifacts to be regarded as benchmarks.

A solution could be provided by existing comprehensive benchmarks that are not yet used in literature \cite{conf/wosp/ReniersLRJ17}.
The YCSB program suite \cite{conf/cloud/CooperSTRS10} offers capabilities for benchmarking key-value and document stores.
For graph stores there are the LDBC Graphalytics Benchmark, LDBC Social Network Benchmark, and GAP Benchmark Suite. 
The GAP Benchmark Suite and the LDBC Graphalytics Benchmark cover kernels common in graph analytics (\eg BFS or PageRank) while the LDBC Social Network Benchmark covers querying workloads.
Recently, a cross NRDS class benchmark was proposed \cite{conf/tpctc/ZhangLXC18}.
These benchmarks, designed for CPU-based systems initially, could also be used to benchmark new designs on FPGAs with modifications for FPGA-specific problems to make them portable.





\paragraph{Performance models}
As established in \cref{sec:fpga}, FPGAs exhibit unstructured parallelism exacerbating comprehension of performance and algorithm complexity on an abstract performance model level.
In the literature we found different ways (cf. \cref{tab:implementations}) to model performance of the proposed solutions breaking down to modeling pipeline and data parallelism in the face of constrained resources (logic resources and memory bandwidth).
This means, the system architecture is broken down into components with known performance (\ie pipeline steps or replicated PEs) where performance is measured in throughput of data which scales linearly with pipeline steps and data parallelism.
Sometimes this is embedded into a roofline model where the performance is first capped by the amount of parallelism and later by the available memory bandwidth.

\paragraph{Summary -- justification}
Justification in the form of workloads and datasets performing well on FPGAs is provided by the literature but the performance measurement landscape is scattered and sometimes specifically tuned to the contribution of the article.
The biggest challenge, however, is the lacking portability of current benchmarks because measurements are performed on vastly different FPGA setups without accounting for \eg different memory bandwidths.
It is sometimes unclear if performance improvements stem from better design or just better hardware.
Thus, we conclude with the following insight:
\begin{insight}
    A portable, relevant benchmark suite that covers all necessary artifacts is missing for robust justification of accelerator usage decisions.
\end{insight}

\subsection{Discussion -- insights}
Over the course of this section, we gained five insights into building an FPGA-accelerated NRDS answering questions Q1--4.
From the motivations of the different NRDS classes in \cref{sec:task} and resulting tasks we saw that two of the biggest challenges of current CPU-based systems are data movement and memory access.
With the patterns we found for placement and memory access optimization, we guide the practitioner towards addressing these challenges with FPGAs regardless of NRDS class augmented with common operator switching strategies.
However, performance largely depends on specific algorithms and data structures designed on a use case basis.
In this regard, we were not able to uncover even more inter-class structure on such a high abstraction level.
Nonetheless, the insights we gained help the practitioner to apply FPGAs to existing or newly designed NRDS, regardless of data model, to take pressure off the CPU or eliminate it from the system architecture completely.
These common patterns might even foster building FPGA accelerators supporting multiple data models at once.

\section{Open research challenges}
\label{sec:challenges}

The literature analysis did not only summarize many interesting solutions but showed several gaps lacking solutions (cf. \cref{tab:implementations}). 
In this section we discuss the important open challenges and research questions we think should be pursued in the near future.

Regarding the differences between the NRDS classes the open challenges vary. 
We found that key-value systems are exceptional in that there already exist complete NRDS (research and commercial).
However, key-value systems are the simplest NRDS class, and their system design leaves many questions unanswered that come up in other classes.
Wide-column stores are not represented in the literature, but solutions from the key-value class should be applicable (cf. \cite{journals/csur/DavoudianCL18}).

The literature on document and graph stores as well as existing accelerator prototypes indicate feasibility.
However, the found solutions in the literature are rather HPC-specific and cannot directly be applied to NRDS.
Thus, there are in general many challenges to be solved towards a complete NRDS.

Besides these rather broad considerations, we identified several open research challenges in the course of this survey that we discuss subsequently.


\paragraph{Non-functional NRDS aspects}
As a broad trend in the literature, the coverage of non-functional NRDS aspects are an open challenge. 
While consistency protocols may transferred from the non-accelerated NRDS literature, it is broadly unclear how to provide production-grade scalability, availability, multi-tenancy, and security with an FPGA-accelerated NRDS.
FPGAs as a relatively new processor architecture for data processing are not integrated as deeply into current systems and do, in contrast to CPUs, not have widely used operating systems providing basic functionality.
Possibly some solutions can be transferred from FPGA-accelerated relational database systems.

\paragraph{Flexibility -- dynamic queries}
The static implementations presented in the current literature need to become more flexible.
There are elegant solutions (\eg skeleton automata \cite{conf/sigmod/TeubnerWN12}) to be found for instructable or parametrizable accelerators that can process more than one rigid operator.
This follows the proposal of domain-specific architectures in \cite{journals/cacm/HennessyP19} and will also largely improve the projected ratio of the workload processed by the accelerator leading to better overall performance of the system.

\paragraph{Comparability -- standard benchmarks}
As discussed in \cref{sec:justification}, there are no commonly used benchmarks in the NRDS classes on FPGAs yet.
A standardized benchmark will be instrumental in gaining more credibility in performance claims, comparability, and justification of FPGAs as an NRDS accelerator.
Moreover, better performance measures help uncover performance impediments in other domains like DRAM (bank parallelism utilization analyzed in \cite{journals/corr/abs-1902-07609}).

\paragraph{Collaborative memory usage}
Data movement overhead dominates accelerated systems and narrows their potential for improvements.
The emergence of cache coherent attachments of FPGAs to the system main memory might alleviate this.
FPGA-directed data movement could take pressure off the CPU and also make more fine-grained acceleration possible.
However, we did not find any solution in the literature for the collaborative usage of the system main memory and smart movement of data.

\paragraph{Near-data memory usage for graph systems}
Especially for document stores, we found deployments of the FPGA in the datapath between CPU and memory or disk (\ie near-memory) to reduce the volume of data being moved to the CPU.
Since graph workloads are highly memory-bound, we see big potential for a tighter integration of FPGA and memory to hide inefficiencies of irregular memory access patterns.

\paragraph{Cross data model processing}
A relatively new trend in NRDS are cross data model systems, \ie systems that allow storing and accessing data in multiple data models simultaneously \cite{journals/csur/LuH19}. 
One example is OrientDB that supports polymorphic queries over graph and document data in one unified system.
FPGAs could, \eg be used near-data to transform and change the binary representation on-the-fly as data is loaded to the CPU.

\paragraph{Cross-accelerator architecture}
As GPUs and FPGAs get more popular and ever more present in the data center, there might be more performance gains in heterogeneous systems using both accelerator types at the same time.
There are first works on those systems in other research areas (\eg \cite{conf/arcs/HampelPM12}), but none in the NRDS literature.
This kind of acceleration might be especially beneficial to NRDS supporting multiple data models, where different workloads are particularly well-suited to different processor architectures.

\section{Summary}
\label{sec:summary}

FPGAs are an instrumental tool in achieving performance gains in data-centric systems in the near future.
In this survey we open up the field of FPGA-accelerated non-relational database systems (NRDS) by studying and answering three hypotheses formulated in \cref{sub:methodology}.

To start with, for hypothesis (H1), \ie \emph{existing non-relational database systems do not realize the potential of FPGA acceleration}, we conducted a system review of commercial NRDS.
We found three experimental extensions to existing systems using FPGAs as accelerators \cite{capiSnap, redisFpga, cassandraFpga}, showing the NRDS-acceleration feasibility but no mainstream adoptions of FPGAs in NRDS, yet.
Hence, we confirm the potential of FPGAs as accelerators for NRDS, but also conclude that this potential is not yet realized in commercial systems.

To give an answer to hypothesis (H2), \ie \emph{there are significant gaps in current research on non-relational FPGA acceleration}, we derive a system aspect taxonomy 
that guides an extensive literature analysis by categorizing the large body of research, for which we provide an insightful overview 
of existing contributions. 

Taking the results of the literature analysis as a knowledge base, we derived common patterns
, confirming hypothesis (H3), \ie \emph{there are patterns guiding the design of an FPGA-accelerated non-relational database system}.
Therefore, we compile a list of four relevant questions that practitioners, like system architects, have to ask themselves when designing and constructing an FPGA-accelerated NRDS.
This survey gives answers to these questions by easy-to-apply patterns for FPGA task definition, FPGA placement, accelerator design considerations, and benchmarking.

In summary, we provide a comprehensive introduction of FPGA acceleration and CPU offload potential for NRDS and present it in a form suitable to everybody interested in the field.
However, we especially regard this survey as a guide for system architects in their decision making and a reference for researchers to guide and conduct new research.

\section*{Acknowledgements} We thank Norman May and Wolfgang Lehner for various valuable discussions in the context of this article.

\bibliographystyle{spmpsci}
\bibliography{fpga_nosql}

\end{document}